\documentclass{spie}
\usepackage{aas_macros}
\usepackage{sidecap}
\usepackage{graphbox}
\usepackage{floatrow}
\usepackage{amsmath,amssymb,array}
\usepackage{hyperref}
\usepackage[]{footmisc}
\usepackage{tabularx}
\newcolumntype{C}{>{$}X<{$}}
\newcolumntype{Y}{>{\centering\arraybackslash}X}

\usepackage{enumerate}
\usepackage{enumitem}

\newlist{considlist}{enumerate}{1}
\setlist[considlist,1]{%
  font=\bfseries Consideration\space ,
  label={\arabic*},
  align=parleft ,
  align=left, leftmargin=0pt, labelindent=\parindent,
  listparindent=\parindent, labelwidth=0pt, itemindent=!
}
\graphicspath{{images/}}

\begin{document}

\title{Variable-delay Polarization Modulators \protect\\ for the CLASS Telescopes}
\author[a]{Kathleen Harrington}
\author[a]{Joseph Eimer}
\author[b]{David T. Chuss}
\author[a]{Matthew Petroff}
\author[a]{Joseph Cleary}
\author[b]{Martin DeGeorge}
\author[a,c]{Theodore W. Grunberg}
\author[a,d]{Aamir Ali}
\author[a]{John W. Appel}
\author[a]{Charles L. Bennett}
\author[a]{Michael Brewer}
\author[e]{Ricardo Bustos}
\author[a]{Manwei Chan}
\author[a]{Jullianna Couto}
\author[a]{Sumit Dahal}
\author[f]{Kevin Denis}
\author[g]{Rolando D\"{u}nner}
\author[f]{Thomas Essinger-Hileman}
\author[g]{Pedro Fluxa}
\author[h]{Mark Halpern}
\author[i]{Gene Hilton}
\author[h]{Gary F. Hinshaw}
\author[i]{Johannes Hubmayr}
\author[a]{Jeffrey Iuliano}
\author[a]{John Karakla}
\author[a]{Tobias Marriage}
\author[j]{Jeffrey McMahon}
\author[a]{Nathan J. Miller}
\author[a]{Carolina Nu\~{n}ez}
\author[a]{Ivan L. Padilla}
\author[k]{Gonzalo Palma}
\author[a,l]{Lucas Parker}
\author[k]{Bastian Pradenas Marquez}
\author[m]{Rodrigo Reeves}
\author[i]{Carl Reintsema}
\author[f]{Karwan Rostem}
\author[a]{Deniz Augusto Nunes Valle}
\author[a]{Trevor Van Engelhoven}
\author[a]{Bingjie Wang}
\author[a]{Qinan Wang}
\author[a]{Duncan Watts}
\author[a]{Janet Weiland}
\author[f]{Edward Wollack}
\author[a,n]{Zhilei Xu}
\author[h]{Ziang Yan}
\author[o]{Lingzhen Zeng}
\affil[a]{Department of Physics and Astronomy, 
		Johns Hopkins University, Baltimore, MD 21218, USA}
\affil[b]{Department of Physics, Villanova University, Villanova, PA 19085, USA }
\affil[c]{Department of Electrical Engineering and Computer Science, Massachusetts Institute of Technology, Cambridge, MA, 02139, USA}
\affil[d]{Department of Physics, University of California, Berkeley, CA 94720, USA }
\affil[e]{Facultad de Ingenier\'{i}a, Universidad Cat\'{o}lica de la Sant\'{i}sima Concepci\'{o}n, \protect\\ Alonso de Ribera 2850, Concepci\'{o}n, Chile }
\affil[f]{NASA Goddard Space Flight Center, Greenbelt, MD 20771, USA }
\affil[g]{
Instituto de Astrof\'{i}sica and Centro de Astro-Ingenier\'{i}a, Facultad de F\'{i}sica, Pontificia Universidad Cat\'{o}lica de Chile, 7820436
Macul, Santiago, Chile }
\affil[h]{Department of Physics and Astronomy, University of British Columbia, \protect\\ Vancouver, BC, V6T 1z4, Canada}
\affil[i]{National Institute of Standards and Technology, Boulder, CO 80305, USA}
\affil[j]{Department of Physics, University of Michigan, Ann Arbor, MI, 48109, USA}
\affil[k]{Departmento de F\'{i}sica, FCFM, Universidad de Chile, Blanco Encalada 2008, Santiago, Chile}
\affil[l]{Space and Remote Sensing, MS D436, Los Alamos National Laboratory,\protect\\
Los Alamos, NM 87544, USA}
\affil[m]{Departamento de Astronom\'{i}a, Universidad de Concepci\'{o}n, Casilla 160 C, Concepci\'{o}n, Chile}
\affil[n]{Department of Physics and Astronomy, University of Pennsylvania, \protect\\ Philadelphia, PA 19104, USA}
\affil[o]{Harvard-Smithsonian Center for Astrophysics, Cambridge, MA 02138, USA }

\authorinfo{Further author information: (Send correspondence to K. Harrington)\\K. Harrington.: E-mail: kharrington@jhu.edu} 

\maketitle

\begin{abstract}
The search for inflationary primordial gravitational waves and the measurement of the optical depth to reionization, both through their imprint on the large angular scale correlations in the polarization of the cosmic microwave background (CMB), has created the need for high sensitivity measurements of polarization across large fractions of the sky at millimeter wavelengths. These measurements are subject to instrumental and atmospheric $1/f$ noise, which has motivated the development of polarization modulators to facilitate the rejection of these large systematic effects. 

Variable-delay polarization modulators (VPMs) are used in the Cosmology Large Angular Scale Surveyor (CLASS) telescopes as the first element in the optical chain to rapidly modulate the incoming polarization. VPMs consist of a linearly polarizing wire grid in front of a movable flat mirror. Varying the distance between the grid and the mirror produces a changing phase shift between polarization states parallel and perpendicular to the grid which modulates Stokes U (linear polarization at $45^\circ$) and Stokes V (circular polarization). The CLASS telescopes have VPMs as the first optical element from the sky; this simultaneously allows a lock-in style polarization measurement and the separation of sky polarization from any instrumental polarization further along in the optical path.

The CLASS VPM wire grids use 50~$\mu$m copper-plated tungsten wire with a 160~$\mu$m spacing across a 60~cm clear aperture. The mirror is mounted on a flexure system with one degree of translational freedom, enabling the required mirror motion while maintaining excellent parallelism with respect to the wire grid. The wire grids and mirrors are held parallel to each other to better than 80~$\mu$m, and the wire grids have RMS flatness errors below 50~$\mu$m across the 60~cm aperture. The Q-band CLASS VPM was the first VPM to begin observing the CMB full time, starting in the Spring of 2016. The first W-band CLASS VPM was installed in the Spring of 2018.
\end{abstract}

\section{Introduction}

The polarized large angular scales of the Cosmic Microwave Background (CMB) are imprinted with signatures from reionization and inflation. Since linear polarization is a spin-2 field, mapping it onto the celestial sphere is most completely accomplished through decomposition into E-modes, a curl free polarization pattern, and the B-modes, a divergence free polarization pattern. The amplitude of the CMB E-mode power spectrum on the largest angular scales ($\ell \lesssim 20$) encodes a measurement of the optical depth to reionization\cite{Watts2018}. The amplitude of the CMB B-mode power spectrum on angular scales up to $\ell \sim 100$ is expected to contain information about the amplitude of the predicted inflationary gravitational waves through the tensor-to-scalar ratio $r$. A detection of these primordial gravitational waves is considered ``smoking gun'' evidence of inflation\cite{national2010New} and is one of the next major goals in cosmology.

Observing the CMB polarization at the largest angular scales requires a high sensitivity measurement of the polarization at multiple frequencies and across large fractions of sky. This necessitates instrument stability and systematic rejection at unprecedented levels. The Cosmology Large Angular Scale Surveyor (CLASS) is a telescope array, sited in the Chilean Atacama Desert, that is uniquely suited to recovering the largest angular scales of the CMB polarization from the ground\cite{Essinger2014, Harrington2016}. The CLASS telescopes cover four frequency bands, one telescope at 40~GHz (Q-band), two at 90~GHz (W-band), and one high frequency dichroic at 150/220~GHz, that span the galactic foreground minimum and enable foreground cleaning at the level necessary to extract cosmological parameters from the CLASS maps\cite{Watts2015}. Each CLASS telescope has a $\sim20^\circ$ field of view and executes constant elevation scans with daily boresight rotations to map $\sim70\%$ of the sky every day. Daily maps and boresight rotations provide a large variety of null tests and systematics checks which are used to track instrument stability and verify the content of the final maps.

\begin{figure}
\centering
\begin{tabular}{cc}
\includegraphics[width=0.38\textwidth , align=c]{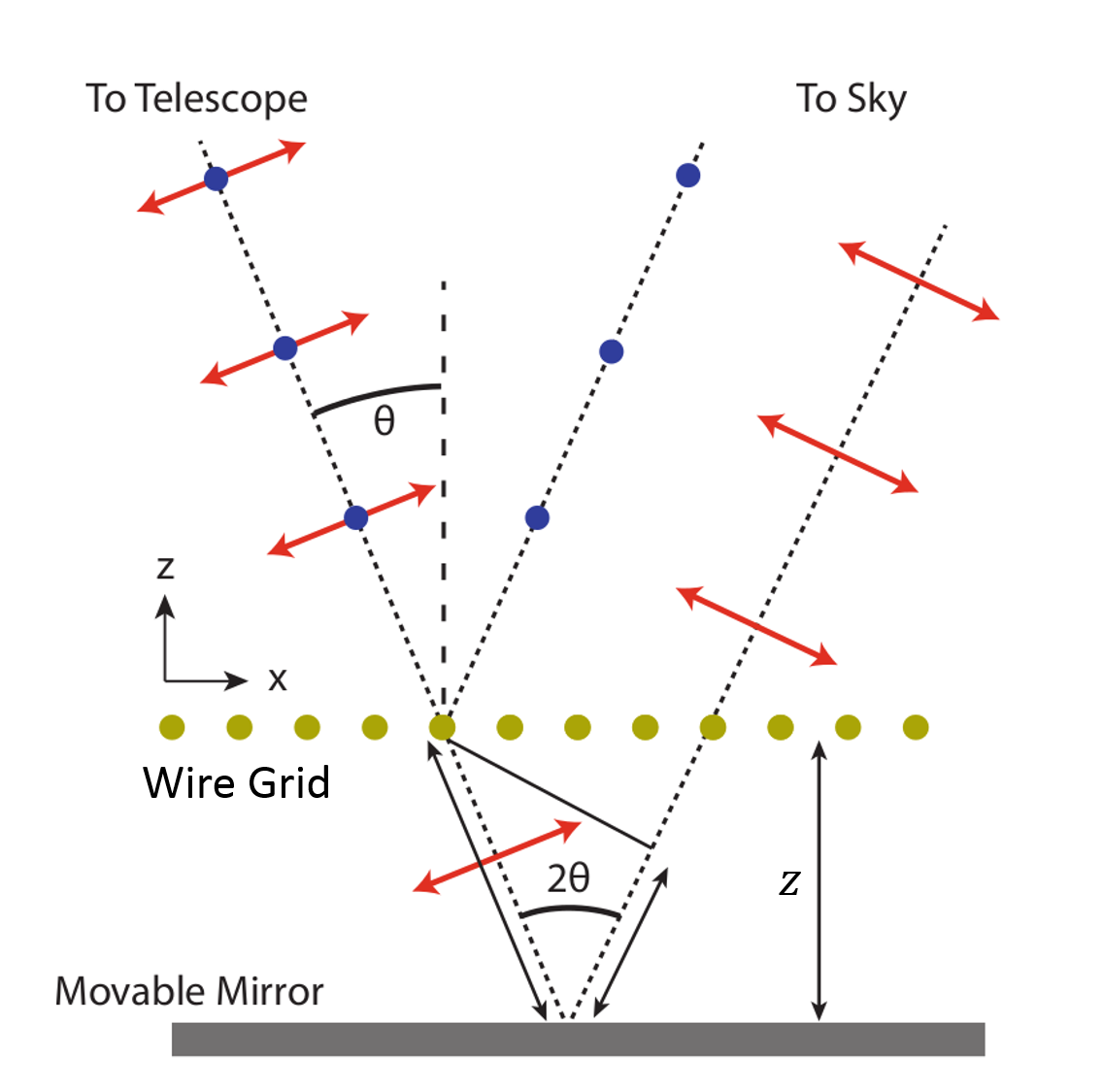}
\includegraphics[width=0.57\textwidth , align=c]{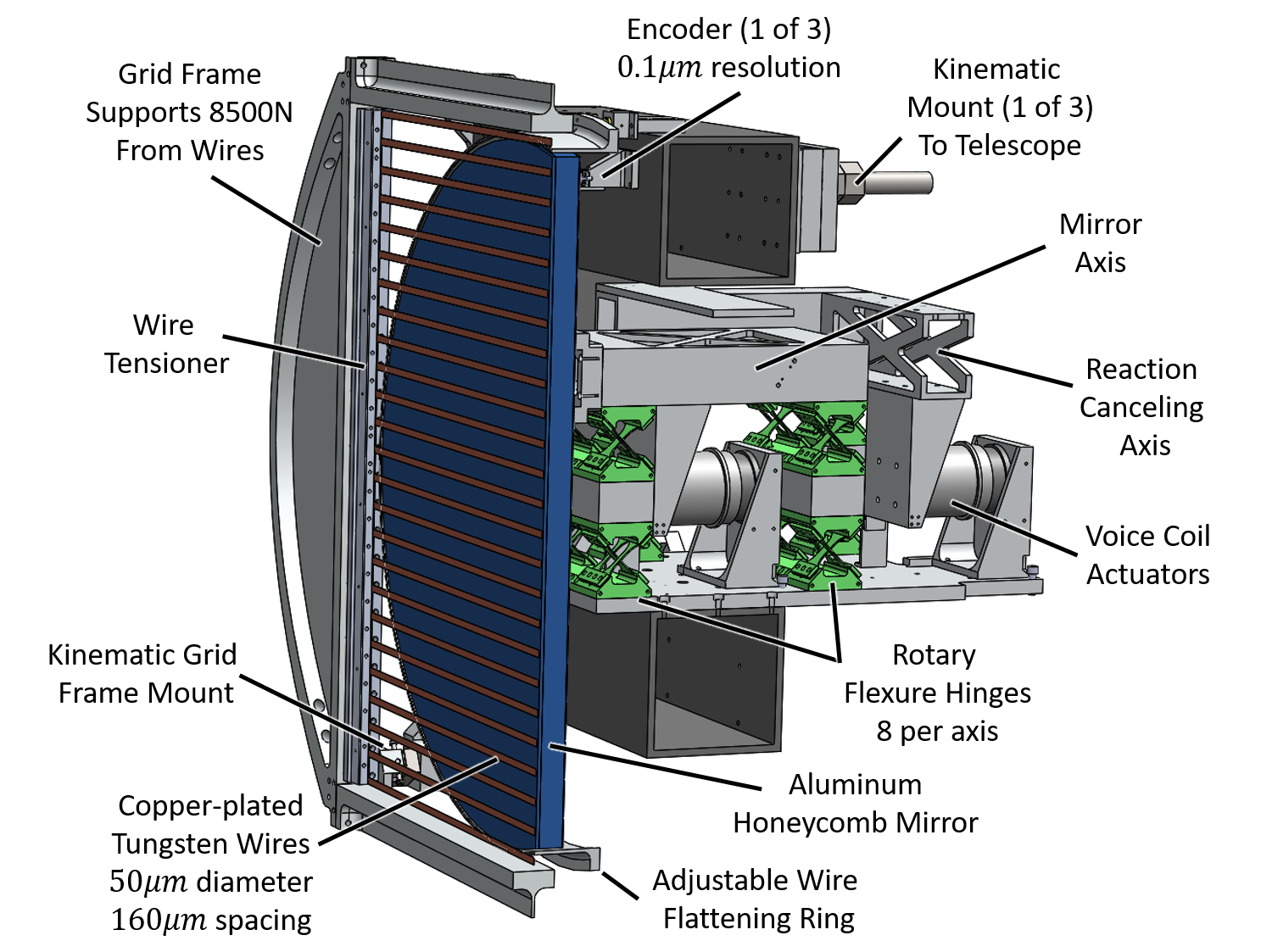}
\end{tabular}
\caption{\label{fig:vpm_overview} (Left) A schematic drawing of the a variable-delay polarization modulator as implemented by the CLASS telescopes. A stationary wire grid is held in front of a movable mirror to induce a phase delay between orthogonal linear polarizations. (Right) A section view of the complete CLASS VPM design.}
\end{figure}

The most unique aspect of the CLASS instrument design is a variable-delay polarization modulator (VPM) in each telescope as the first optical element from the sky. Shown in figure \ref{fig:vpm_overview}, VPMs use a linearly polarizing wire grid in front of a moveable mirror to induce a variable phase delay between polarization states parallel and perpendicular to the direction of the wires in the grid. This phase delay modulates one of the linear polarization states (Stokes $Q$ or $U$) with circular polarization (Stokes $V$) as a function of the distance between the wire grid and the mirror. Setting the modulation at a rate faster than the rate of change of the sky or instrument creates a lock-in style measurement of the polarization at a particular point on the sky and greatly increases the stability of the demodulated data. In addition, instrument polarization ($T\rightarrow P$) after the VPM is unmodulated and does not affect the demodulated data. Since the VPM is the first optical element from the sky, only modulated polarized ground pickup or instrument polarization from the telescope baffling or the VPM can influence the demodulated data. This significantly mitigates the level of $T\rightarrow P$ leakage observed in the CLASS telescopes. The combination of instrument stability and systematics rejection due to the front-end polarization modulation with a VPM enables the recovery of the largest angular scales of the CMB polarization from a ground based telescope\cite{Miller2016}.

These proceedings cover the requirements, design, implementation, and performance of the VPMs for the first two CLASS telescopes (Q-band and the first W-band telescope). The Q-band VPM was installed at the CLASS site in the Spring of 2016 and was on sky until the Spring of 2018, when it was removed and re-installed. The first W-band VPM was installed in Spring 2018.

\section{VPM design requirements}

The scientific goals of the CLASS survey drive the design requirements for the CLASS VPMs. The B-mode signal from primordial gravitational waves is predicted to be higher than the gravitationally lenses E-modes on angular scales larger than about a degree on the sky; this sets the upper limit for the CLASS beamwidths. Since the VPM is the first optical element, the diameter of the VPM limits the beamwidth of the CLASS telescopes. The diffraction limit, $\theta \sim \lambda/d$, and the Q-Band center wavelength of 7.89~mm indicate that the CLASS VPMs must have a clear aperture $\sim 45$~cm. The CLASS VPMs have been constructed with a 60~cm clear aperture to limit spill and reduce systematic effects resulting from edge illumination of the VPM\cite{Eimer2012}.

A non-normal incidence is required since the VPM is a reflective device and normal incidence would require an obstruction, causing an unacceptable level of diffraction and instrument polarization in the system. However, non-normal incidence on a flat mirror produces a polarized signal that increases with angle of incidence, so it is desirable to keep the angle of incidence on the VPM as small as possible. CLASS has chosen to set the angle of incidence at $20^\circ$ as a compromise between these two design constraints.

CLASS implements a polarization modulator to move the scan frequency to a region above the $1/f$ knee of the detectors. The $1/f$ knee frequency depends on the thermal stability of the atmosphere, warm optics, and cryogenic receiver as well as the white noise level from the detectors. In addition, the modulation must be slower than the response rate of the detectors. Given these constraints and the noise levels observed in reference \citenum{Appel2014}, the CLASS modulation frequency is set at 10~Hz.

\subsection{Throw Requirements}

Given the angle of incidence on the VPM and the band-pass of each detector, we can calculate the required throw, the grid-mirror distance range through which the VPM mirror moves, for each telescope. For an ideal VPM with a perfect mirror and grid, the single frequency, $\nu$, phase delay between the two orthogonal linear polarization states is
\begin{equation}
\phi = \frac{4\pi\nu}{c} z\cos{\theta},
\end{equation}
\noindent where $z$ is the distance between the grid and the mirror and $\theta$ is the angle of incidence. This causes mixing between one of the linear polarization states and circular polarization. If $+Q$ is defined as being along the wires of the VPM, then the polarization before and after the VPM is modulated as
\begin{equation}
\begin{split}
U_\mathrm{out}(\nu) & = U_\mathrm{in}(\nu) \cos{\phi} + V_\mathrm{in}(\nu) \sin{\phi}  \\
V_\mathrm{out}(\nu) & = -U_\mathrm{in}(\nu) \sin{\phi} + V_\mathrm{in}(\nu) \cos{\phi}.
\end{split}
\end{equation}

\begin{figure}
\centering
\includegraphics[width=0.9\textwidth]{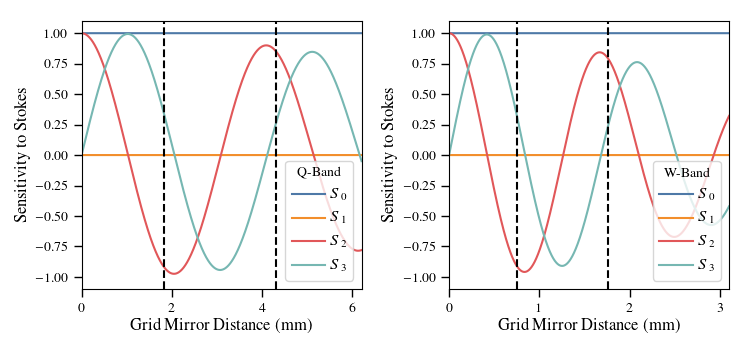} \\
\caption{\label{fig:mod_funct} Modulation functions for ideal CLASS VPMs with a single linearly polarized Q-Band (left) and W-Band (right) detector aligned at $45^\circ$ with respect to the wire grid at an angle of incidence of $20^\circ$. $S_x$ refers to the detector's sensitivity to Stokes $I$, $Q$, $U$, and $V$ respectively. The black dashed lines delineate the optimized mirror throw for each band.}
\end{figure}

\noindent The single frequency modulation is integrated over the band-pass of the detectors to calculate the total modulation function of the instrument. For the following calculations we assume this ideal model for the VPM with a top-hat band-pass for each detector. The band edges for the detectors are 33-43~GHz and 77-108~GHz for the Q-band and W-band telescopes, respectively. The frequency integration also assumes a source spectrum of a 2.725~Kelvin blackbody. The total modulation functions\footnote{The modulation functions are the top row of the Mueller matrix for the combined VPM plus detector system, this takes polarization from the sky and returns the power observed by the detector.} for the CLASS Q-Band and W-band VPMs are shown in figure \ref{fig:mod_funct}, where $S_x$ describes the sensitivity to Stokes $I$, $Q$, $U$, and $V$ as a function of grid-mirror distance for a linearly polarized detector aligned at $45^\circ$ with respect to the wire grid. The sensitivities are normalized to the intensity observed by a single linearly polarized detector. The decrease in the amplitude of the peaks with increasing grid-mirror distance is a decoherence effect due to the finite band-pass of the CLASS detectors.

The VPM modulation produces a detector timestream, $P_\textrm{det}(t)$, that depends on the modulation functions ($S_x$) and the polarization state of the incoming radiation
\begin{equation}
P_\mathrm{det}(t) = S_{0}\left[z(t)\right] I + S_{1}\left[z(t)\right] Q +S_{2}\left[z(t)\right] U + S_{3}\left[z(t)\right] V,
\label{eq:det_tod}
\end{equation}
\noindent where $z(t)$ is the timestream of the grid-mirror distances for a chosen sinusoidal VPM throw.
\begin{figure}[t]
\centering
\begin{tabular}{c}
	\includegraphics[width=0.95\textwidth]{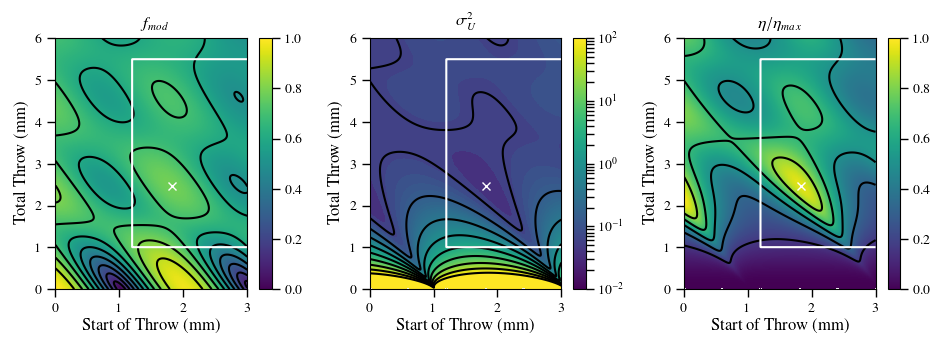} \\
	\includegraphics[width=0.95\textwidth]{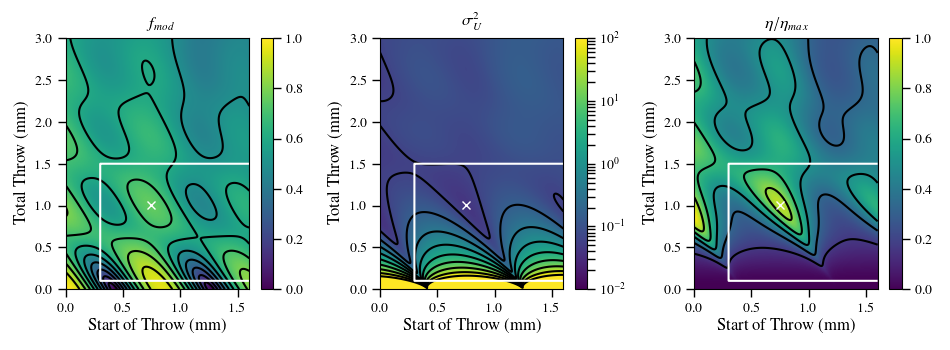} \\
\end{tabular}
\caption{\label{fig:throw_opt} The modulation efficiency (left), uncertainty on $U$ (center), and relative efficiency (right) for ideal CLASS-style VPMs with Q-band (top) and W-band (bottom) detectors over a range of mirror throw parameters. The start of the throw is the minimum distance between the grid and the mirror and the total throw is the peak-to-peak amplitude of the mirror motion. The white boxes indicate the regions accessible to the mechanical/electricial design of the CLASS VPMs, and the white X's mark the optimal throw parameters.}
\end{figure}

For an ideal VPM with a detector aligned at $45^\circ$ with respect to the VPM wires, $S_0$ and $S_1$ are one and zero, respectively. Stokes $V$ (circular polarization) is predicted to be zero for the CMB, so the optimum throw for the CLASS VPMs will maximize the time spent observing $U$ while minimizing the uncertainty $\sigma_U$. The modulation efficiency is the mean sensitivity to Stokes U over the modulation cycle
\begin{equation}
f_\mathrm{mod} = \lim_{T \to \infty} \left( \frac{1}{T} \int^{T}_{0} S_2\left(z(t)\right)^2 dt \right)^{1/2}.
\label{eq:fmod}
\end{equation}

\noindent The uncertainty on $U$ depends on the number of points observed through the throw. For this we assume 40 points through the throw and set up a linear least squares fit where 
\begin{equation}
\vec{d} = M\vec{S}, \quad
M = \begin{bmatrix}
S_{0}\left[z(t)\right] & S_{2}\left[z(t)\right] & S_{3}\left[z(t)\right] \\
\end{bmatrix},
\quad \text{and} \quad
\vec{S} = (M^T M)^{-1} M^T \vec{d}.
\label{eq:tod_fit}
\end{equation}
\noindent Here, $\vec{S}$ is the vector of Stokes parameters which are modulated by the VPM and observed by the detector in equation \ref{eq:det_tod}. The uncertainty on $U$ is calculated using the (2,2) element of the inverse Fisher matrix
\begin{equation}
\sigma_U^2 = (M^T M)^{-1}_{22}.
\end{equation}
\begin{table}
\begin{tabularx}{\linewidth}{|l|Y|Y|Y|Y|}
\hline
  & Throw Start~(mm) & Total Throw~(mm) & Modulation Efficiency~($f_\mathrm{mod}$) & Relative Efficiency~($\eta/\eta_\mathrm{max}$) \\
\hline
Q & 1.828 & 2.4775 & 0.7855 & 0.9561 \\
W & 0.7476 & 1.0107 & 0.7564 & 0.898 \\
\hline
\end{tabularx}
\caption{\label{tab:throw_params} The optimized VPM throw parameters for the Q-Band and W-Band CLASS VPMs. The differences in throw start and total throw are due to the wavelength shift between the two detector bands while the difference modulation efficiency and relative efficiency is entirely due to the difference in relative bandwidth.}
\end{table}
These two metrics are combined into an observation efficiency $\eta=f_\mathrm{mod}/\sigma_U$. The optimal throw for a VPM is where $\eta$ is maximized for the bandpass of the detectors. Figure \ref{fig:throw_opt} shows the modulation efficiency ($f_\mathrm{mod}$), the uncertainty on $U$ ($\sigma_U$), and the relative efficiency for a variety of throw parameters for the Q-Band and W-Band telescopes. The white box indicates the region of the plot accessible by the mechanical and electrical design of the VPMs. The highest peaks in the efficiency plot occur at zero path length difference and are not considered viable due to the fact that they would require contact between the mirror and the grid. Table \ref{tab:throw_params} lists the optimized throw parameters, that are marked by a white X in figure \ref{fig:throw_opt}. The differences in modulation efficiency and relative efficiency between the two bands are a consequence of the different detector bandwidths. 

\subsection{Alignment Requirements}
The VPM mirror must be held extremely parallel to the wire grid throughout the entire throw of the mirror. To zeroth order, the tilt or parallelism error between the mirror and grid must be much less than the full width half max of the CLASS beams; however, it is useful to quantify this requirement to place design constraints on the VPM. A misalignment between the mirror and grid creates a pointing difference between the two surfaces; where a difference in temperature between the two directions leads to an additional term in the intensity modulation function
\begin{equation}
S_\mathrm{0, tilt} = S_0 + \left.\frac{dT}{d\delta}\right\vert_{\delta_0} (\delta(z) - \delta_0)
\label{eq:tilt_TP}
\end{equation}
\noindent where $\delta_0$ is the pointing of the VPM grid and $\delta(z)$ is the pointing of the mirror across the mirror throw. From equation \ref{eq:tilt_TP} it is clear that a constant tilt across the mirror throw will not be modulated by the VPM, meaning only changing tilts will cause a modulated temperature to polarization leakage ($T\rightarrow P$). 

For a systematic modulated signal to influence the demodulated data, it must create a bias in the Stokes parameters fit using equation \ref{eq:tod_fit}. The largest source of temperature fluctuations will be the atmosphere where, due to effective thickness variations as a function of elevation, a 10~K sky changes by $\sim69$~$\mu$K/arcsecond at CLASS pointing of $45^\circ$. Simulating a timestream with equation \ref{eq:tilt_TP} and fitting the output Stokes parameters with equation \ref{eq:tod_fit} gives $\Delta U = -30.5$~$\mu$K and $\Delta Q=-0.4$~$\mu$K per arcsecond of mirror tilt. This is a $T\rightarrow P$ leakage of $3\times 10^{-6}$ per arcsecond. To maintain a $T\rightarrow P$ leakage of less that $10^{-4}$, the change in tilt across the mirror throw must be less than 30~arcseconds, and the lower the tilt is across the throw the lower the $T\rightarrow P$. The requirement of a changing mirror tilt of less than 30~arcseconds must hold for all boresight rotations the CLASS telescope completes.

The realized $T\rightarrow P$ leakage will be less than described above because the effect can be reduced through various data analysis techniques. For example, the $S_0$ term used in the demodulation can be replaced with equation \ref{eq:tilt_TP}, or two detectors can be pair subtracted before demodulation to reduce the effect to the level of difference in the two beams. 

\section{Mirror Transport Mechanism}

From the previous section, CLASS requires a VPM with a 60~cm mirror able to move through a range of up to 2.5~mm along the vector normal to its surface, at 10~Hz, with less than 30~arcseconds of variable tilt across the mirror throw. This mirror must be mounted $\sim 1$~mm behind a fixed linearly polarizing wire grid which also requires a clear aperture of at least 60~cm. Both surfaces must be held at an angle of incidence of $20^\circ$ ($25^\circ$ from vertical when observing at $45^\circ$ elevation). These parameters must be maintained throughout $\pm45^\circ$ boresight rotations that change the direction of gravity by $\sim20\%$ at the VPM. This section describes the design and control of the Mirror Transport Mechanism (MTM) used to move the mirror, and section \ref{sec:wiregrid} describes the design and manufacturing of the wire polarizer.

\subsection{\label{subsec:mech_deisgn} Mechanical Design}

Flexures are compliant mechanisms that use flexing or deflection to accomplish a desired outcome \cite{compliant_mechanisms}. There are countless types of flexures with one or more translational or rotational degrees of freedom where the dimensions, configuration, and materials can be engineered to achieve the exact level of compliance or constraint desired in every direction of motion. One key aspect of a flexural system is the motion and motion constraints are achieved with no contact or friction between parts. Assuming the forces on the flexure do not exceed the yield strength of the material, this enables extremely high wear resistance and repeatability over long-lifetimes.

\begin{figure}
\centering
\begin{tabular}{ccc}
	\includegraphics[height=1.75in]{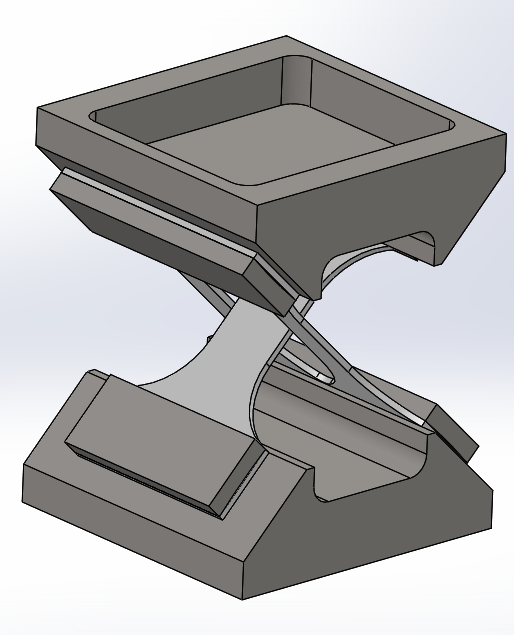} &
	\includegraphics[height=1.75in]{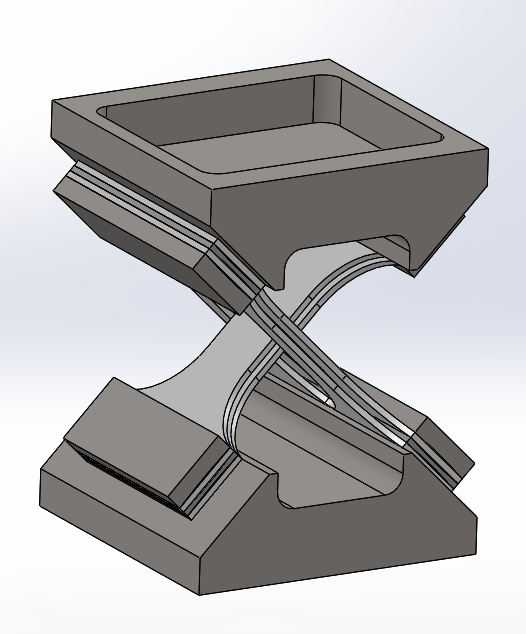} &
	\includegraphics[height=1.75in]{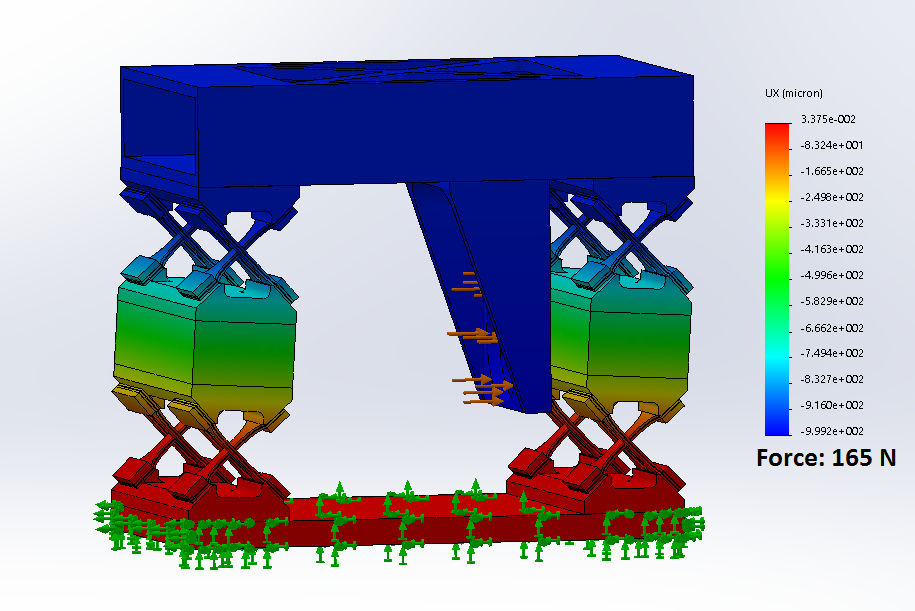}\\
 	Q - Band & W-band & Four bar linkage Simulation
\end{tabular}
\caption{\label{fig:flexures} The crossed leaf rotary flexures used for the Q-band (left) and W-band (center) MTMs. The analytic description of these devices is in section \ref{subsec:mech_deisgn} while the explanation of the differences is in section \ref{subsec:W_redesign}. The right image shows the 3D four-bar-linkage configuration used to connect the CLRFs under an example deflection simulation that was used to estimate the final mechanical properties of the MTM.}
\end{figure}

The design requirements for VPM Mirror Transport Mechanisms (MTMs) are very well matched to a flexural mechanism with one translational degree of freedom. Previous VPMs designed for Hertz \cite{Krejny2008} and PIPER \cite{Chuss2014} both used double-bladed flexures to guide mirror motion. However, the larger total required throw and desired lifetime of $>5$~years for the CLASS VPMs produced addition challenges and required a more complex flexure system.

CLASS VPMs use ``cross-leaf rotary flexures''(CLRF) \cite{website:Baltec} as a basic building block for the MTMs. These CLRFs are shown in the left and center images in figure \ref{fig:flexures} and use the deflection of two crossed leaf springs to create a rotation around a pivot axis located where the springs cross. Basic leaf springs have a rectangular cross-section ($w \times t$ - width $\times$ thickness) with second moments of area of $w t^3/12$ and $w^3 t/12$ when the force is along the thinner and wider directions, respectively. Since the second moment of area scales with the cubed thickness of the axis of deflection, setting $w >> t$ creates a spring with one rotational axis much stiffer than the other. As shown in figures \ref{fig:vpm_overview} and \ref{fig:flexures}, the CLRFs are built into a parallelogram four-bar-linkage configuration where they are separated by a vertical distance, $h$, to translate rotations around the CLRF axes into horizontal displacement. The spring constant of the complete MTM is 
\begin{equation}
k_\mathrm{x, MTM} = \frac{\sqrt{2} Ewt^3}{3 h L^2} \text{ N/m}.
\label{eq:kmtm}
\end{equation}
\noindent where $E$ is the elastic modulus of the spring material and $L$ is the length of the crossed leaf springs. A variety of design constraints and considerations were used to determine the dimensions and materials of the MTM flexure:

\begin{considlist}
\item The material selection of the leaf springs sets the value for the elastic modulus in equation \ref{eq:kmtm}. The CLASS MTMs need to survive for 5 or more years of operation while cycling at 10~Hz, requiring over 1.6 billion MTM cycles. To ensure survivability, we use wear-resistant blue tempered spring steel which is specifically tempered for applications such as the MTMs. Spring steel is only commercially available in specific thicknesses and dimensions, so we constrain our designs to use only these available options.

\item The maximum required throw of the MTM, $\sim2.5$~mm, must be well within the linear regime of deflection for the spring steel straps to prevent plastic deformation. In addition, higher repetitive stresses applied to the steel straps result in lower lifetimes. Since CLASS requires a significant number of cycles we need to keep the stresses at a manageable level. For a given set of leaf spring dimensions, the spacing between the axes, $h$, can be increased to reduce the stresses imparted on the leaf springs while maintaining the off-axis rotational constraints.

\item \label{lr:drive_selection} The selection of the axis drive system significantly influences the design of the flexure. The flexure could be driven by any type of linear actuator or a cam-type system coupled to any rotational motor. We chose a linear actuator because cams would require the entirety of the mirror throw to be physically designed into the system, which would prevent adjustments to the throw once the VPM was installed. In addition, the cams would be the only point of contact in the system and were likely to wear down with use while producing unnecessary vibrations in the system. Voice coil linear actuators, solenoids mounted in a magnetic field, are used to drive the MTMs because they provide a non-contact force which is linear to drive current and enable adjustments to the mirror positioning through calibration and commissioning of the instrument. The commercially available voice coils limited the maximum force which could be imparted on the MTM flexure and created an upper bound for the possible flexure stiffness.

\item \label{lr:mirror_weight} The mirror, described in more detail is section \ref{subsec:mirror}, is the main source of mass and drag in the system. For a given maximum force output of the voice coils, the flexure needed to be compliant enough to move the mirror through the entire required mirror throw. This requirement needed to be achieved for all gravitational vectors encountered by the MTM over all elevation and boresight rotations.

\item \label{lr:res_freq} The combination of the mirror mass and MTM spring constant results in a resonance frequency for the system as a whole. Ideally, the resonance frequency is significantly higher than the drive frequency of the system. In practice, other constraints on the system such as the required mirror throw and drive selection push the resonance frequency lower than ideal. This has consequences for the mirror control and vibration suppression which are described in section \ref{subsec:mtm_control}. 
\end{considlist}

The candidate system was simulated using the finite element analysis (FEA) software in Solidworks\textsuperscript{\textregistered}. The right image of figure \ref{fig:flexures} shows the simulated deflection of the MTM mirror axis under a force applied at the voice coil mount that is through the center of mass of the mirror. This simulation was used to obtain a more precise prediction of the spring constant of the system and to verify that the linkages between the CLRFs were stiff enough to prevent any undesired deflections.

In addition, the simulation was used to extract the torsion coefficients for the mirror axis because the required alignment between the mirror and wire-grid creates a rotational constraint to torques for the MTM. For the two rotations of interest, the torsion coefficients were both $\sim20\times10^6$~Nm/rad, which indicates a required torque of $\sim100$~Nm to produce one arcsecond of deflection. Since static deflections are unmodulated, the MTM would require \textit{changing} torques of this order of magnitude to produce these deflections which is well outside any expected forces during VPM operation. In addition, non-uniformities in the CLRFs cause non-uniform deflections which have to propagate through the steel and aluminum connections to influence the mirror tilt. For standard machining and assembly procedures with errors on the order of $\sim50\mu$m, these deflections are small but non-negligible, and great care was taken to minimize assembly non-uniformities wherever possible.

The leaf springs were manufactured using a wire-cut electronic discharge machine (EDM). EDMing was used because it can produce higher precision parts than laser cutting or water jetting ($\sim100$~$\mu$in compared to $\sim1000$~$\mu$in) and causes less heat damage or metal warping during the machining. The leaf spring holders and clamps are CNC milled steel to match the leaf springs while the spacers between the CLRFs are aluminum tube stock, selected to maintain flexure stiffness while reducing the weight of the system. Each MTM flexure was assembled using alignment jigs and pins for the CLRFs to maintain uniformity between them at a level of 25~$\mu$m or better. The voice coil drives also required alignment jigs to achieve uniformity in the motor constants since misalignments create nonlinearities in the current to force ratio.

The mirror positions are read out using 0.1~$\mu$m resolution linear optical  quadrature digital encoders\footnote{Renishaw ATOM\textsuperscript{TM} incremental encoder system with RTLF linear scale.}. Optical encoders were chosen because they are non-contact and have sufficient resolution for the mirror displacements required for the MTMs. In addition, quadrature digital encoders have a standard interface which allows for significant flexibility in controller hardware. Three encoders (see figure \ref{fig:vpm_overview}) are mounted on the back of the mirror to enable tip/tilt measurements of the mirror in addition to the mirror position. The encoders and tapes must be mounted parallel to the direction of motion of the mirror to prevent bias in the position readout.

The deflection of MTMs were measured using the voice coils to apply forces and the encoders to read out position; an example of these measurements is shown in figure \ref{fig:q_v_w_control}. The prediction of the MTM spring constant (equation \ref{eq:kmtm}) matches the final measured values to within $6\%$, validating that our simple MTM spring model is well motivated for understanding the trade-offs and design constraints in the system. The simulation calculated spring constant was within $4\%$ of the measured value. Some of the difference between the calculated, simulated, and measured values is likely due to the calibration of the motor constant of the voice coil, which is not well constrained.

\subsection{\label{subsec:mirror}Mirror Selection and Mounting}

The mirror is the largest source of mass and drag for the MTM system and therefore has important mechanical and dynamic effects in addition to its function as an optical element in the VPM. Optically, the mirror needs to be a flat mirror with low surface roughness, high reflectivity, and low emissivity. The CLASS VPM mirrors are constructed using aluminum sheet metal as the mirror face. Aluminum is chosen because of its manufacturability, durability, and high conductivity. We opted not to coat the mirror with copper or gold because the increased conductivity would have resulted in an improvement in emissivity of less than 3~mdB. The surface roughness of the rolled aluminum (8-16~$\mu$in) is sufficient to maintain the phase across the illuminated surface and produces an acceptable emissivity of less than 0.2~mdB. Hence, no further processing of the surface is required.

Aluminum sheet metal alone is not stiff enough to maintain its shape under the oscillatory forces subjected to the VPM mirror, so a honeycomb panel is used to significantly increase the mirror stiffness while minimizing the addition of weight. Various honeycomb panel manufacturing methods were used to determine the method which produced the best mirror. The mirror installed in the Q-band VPM was custom made using a 6.35~mm thick precision aluminum plate as a face sheet. The face sheet, honeycomb, and thinner back sheet were individually cut to a 61~cm diameter and then pressed into a panel. This fabrication method created a defocus in the VPM mirror which is visible in figure \ref{fig:gm_alignment} and is large enough it would be unacceptable for the higher frequency mirrors. The thick plate as a face sheet was also heavy, causing issues for the control system. For the W-Band mirror, a 4~ft by 8~ft panel with a 1.27~mm thick face sheet was assembled using standard commercial practices and then 61~cm diameter mirrors were cut out of the panel. This method led to mirror faces that have flattness errors well below 25~$\mu$m and a significantly reduced mirror weight. 

Each VPM mirror must be mounted perpendicular to the direction of motion, because misalignments will produce errors in the mirror positions read out by the encoders. Usually, mirrors have kinematic and adjustable mounts; however, the VPM mirrors were permanently epoxied onto the MTM to prevent the large, repetitive vibration forces from causing separation during operation.\footnote{Optical alignment of the VPM to the rest of the telescope optics is performed using kinematic adjustments on the back of the VPM MTM mount, shown in figure \ref{fig:vpm_overview}, where the vibrations of the mirror motion have been canceled by the reaction axis and kinematic mounting is safer.} To align the mirror to the MTM axis, adjustable alignment jigs, set in place using measurements from a portable coordinate measuring machine (CMM), were used to hold the mirror and mount while the epoxy was applied and allowed to cure. The misalignments between both the Q- and W-band mirrors and the MTM axes were at or below the level of systematic error expected from the CMM ($\sim10$~arcseconds).

\subsection{\label{subsec:W_redesign} Differences between Q-Band and W-Band VPMs}

The difference in the VPM mirrors, described in section \ref{subsec:mirror}, is one of the changes made based on lessons learned from the development of the Q-Band VPM. The weight of the mirror and the compliance of the flexures combined to produce a resonance frequency that was too low (consideration \ref{lr:res_freq}, more details in section \ref{subsec:mtm_control}). While it was possible to create a working control system for Q-band under these conditions, it left the MTM sufficiently susceptible to external disturbances that it would have been problematic for the higher frequency telescopes where the minimum distance between the mirror and the grid (table \ref{tab:throw_params}) is significantly decreased. 

\begin{figure}
\centering
\begin{tabular}{cc}
\includegraphics[align=c, width=0.48\textwidth]{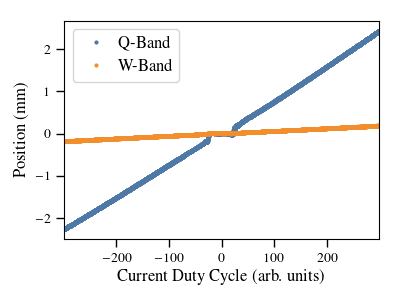} & 
\includegraphics[align=c, width=0.4\textwidth]{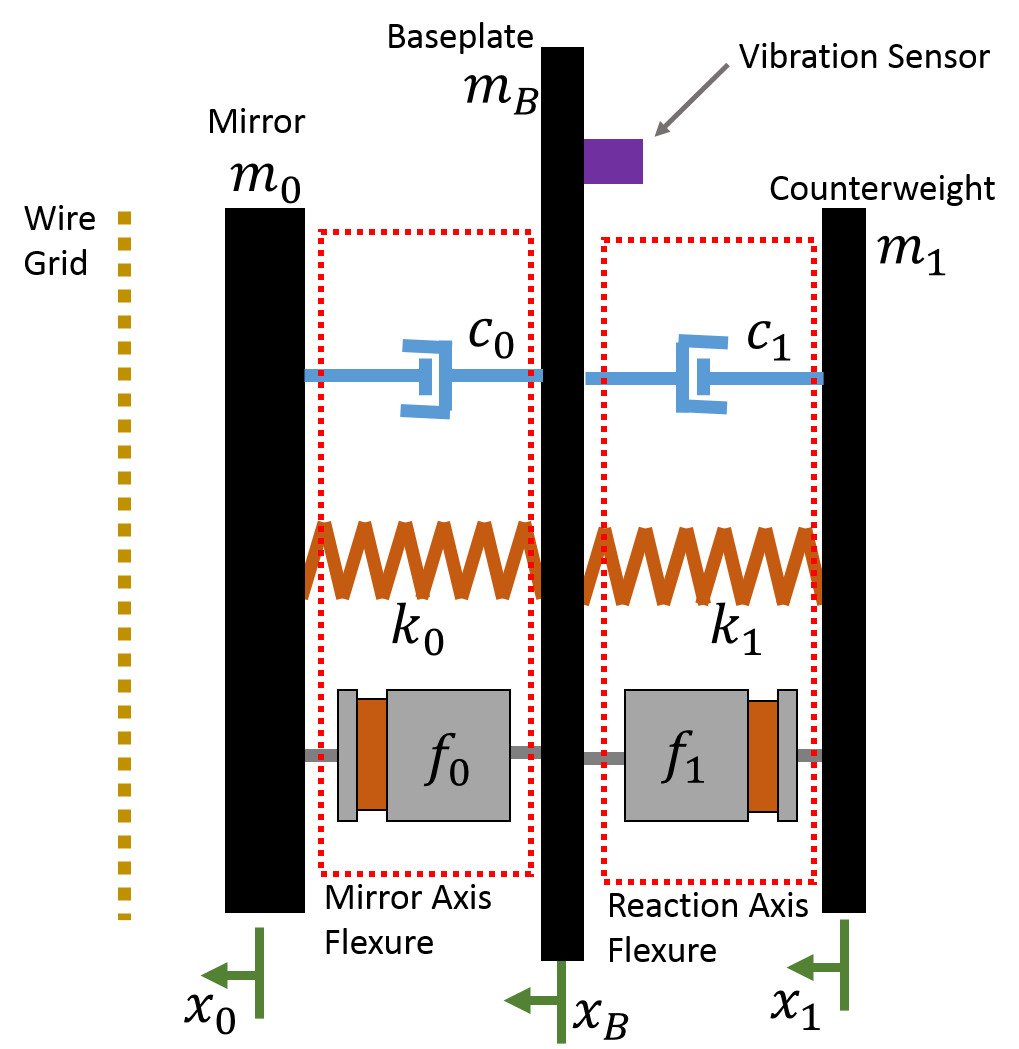} \\
\end{tabular}
\caption{\label{fig:q_v_w_control} (Left) The MTM deflection as a function of current duty cycle (approximately equal to force) for the Q-band and W-band VPMs illustrates the factor of 12.5 increase in stiffness between the two designs. The Q-band curve also shows the dead-zone in the motor driver due to the large inductance of the voice coils interacting with the pulse width modulation in the driver. (Right) The mechanical equivalent diagram for the MTM flexure, which functions as a coupled damped mass spring system where the spring constants, $k$, are from the MTM flexures and the damping, $c$, and mass, $m$ are primarily from the mirror and mirror counterweight.}
\end{figure}

In order to minimize the differences between the VPMs, methods of altering the CLRFs were investigated. The most straightforward way to increase the spring constant of the CLRFs was to increase the thickness of the leaf springs, but the required increase was large enough that blue-tempered spring steel is not available in those thicknesses. Instead, as shown in figure \ref{fig:flexures}, a second leaf spring was added to each original spring with a small spacer in between. It was found that the thickness of the spacer could be chosen to finely adjust the resulting spring constant and resonance frequency. The W-band MTM was made 12.5 times stiffer than the Q-band MTM. This was as stiff as possible given the available voice coil drives (consideration \ref{lr:drive_selection}) and the weight of the W-band mirror (consideration \ref{lr:mirror_weight}). Figure \ref{fig:q_v_w_control} shows the deflection versus current duty cycle ($\sim$force) for Q and W-band VPMs and illustrates the large increase in stiffness. 

\subsection{\label{subsec:mtm_control} MTM Control}

The control system for the MTM has two design objective: move the mirror through the required throw at 10~Hz and cancel the impulses created by this motion. The impulse canceling is done with a reaction canceling axis (see figure \ref{fig:vpm_overview}) that surrounds the mirror axis and is as identical to it as possible. The reaction axis has a separate voice coil driver and encoder readout. Since both the mirror and reaction axes are constrained to move in only one direction, the small parasitic vertical vibrations created through the MTM rocking are deliberately ignored. Measurements of vibration levels for the fielded VPMs validated this decision as any parasitic vertical vibrations are much smaller then the external vibrational disturbances. 

Since the MTM is a flexure system that is completely non-contact and only experiences small deflections around the zero position, the dynamics are within the linear regime and can be described by linear control theory. The right diagram in figure \ref{fig:q_v_w_control} shows the equivalent mechanical system for the MTMs, which functions as a coupled and damped mass spring system with two force inputs from the voice coils and two outputs from the encoders on the mirror\footnote{There are three encoders mounted on the back of the mirror to enable measurements of tilts across the mirror. Since there is only one possible input to the mirror side, only one encoder is read out into the control system.} and reaction axes. The spring constants, $k$, come from the MTM flexure. The mirror has a mass, $m_0$, and is also responsible for much of the damping, $c_0$, experienced by the mirror axis ($c_0 \sim \pi r^2$). The reaction axis has a counterweight attached to the back of the flexure; this counterweight alters the damping $c_1$ of the reaction axis as well. External disturbances, such as vibrations from the telescope structure, enter the system through the baseplate of the MTM flexure.  A vibration sensor is mounted to the baseplate to track vibrations induced by the MTM as well as any incoming external disturbances.
\begin{figure}[t]
\begin{tabular}{c}
	Q-Band Frequency Response Function \\
	\includegraphics[width=0.85\textwidth]{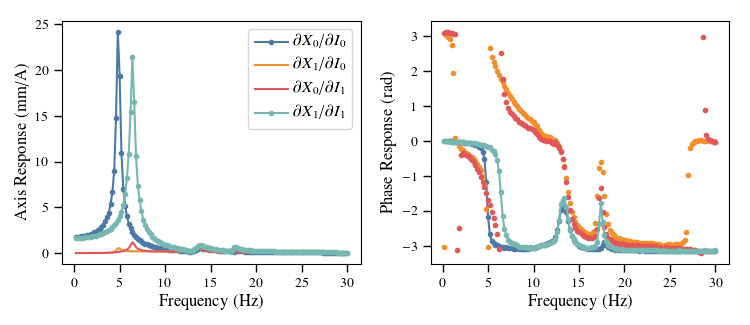} \\
	W-Band Frequency Response Function \\
	\includegraphics[width=0.85\textwidth]{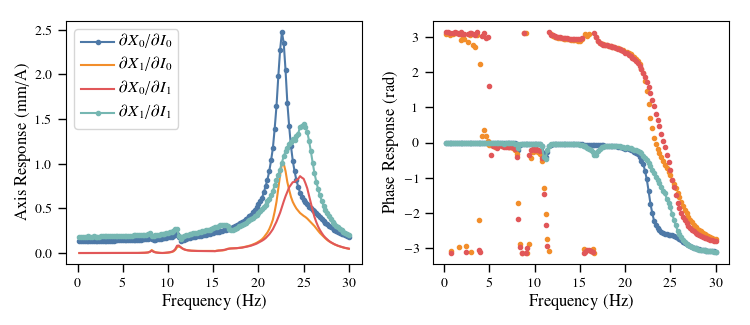} 
\end{tabular}
\caption{\label{fig:frf_meas} The measured frequency response functions of the Q-band (top) and W-band (bottom) MTMs, taken at the CLASS site before the installation of the wire grid. $\partial X_i/\partial I_j$ labels the response of the $i^{th}$ axis to the drive of the $j^{th}$ axis where 0 is the mirror axis and 1 is the reaction axis. The increase in MTM stiffness and decrease in mirror weight for the W-band VPM increased the resonance frequency by a factor of $\sim4.5$ and reduced the resonance height by a factor of $\sim10$. The response of one axis to the drive of the other indicates the axes are not completely independent and that the mounting structures had a finite stiffness. The smaller features in the response are due to the VPM coupling to the mounting utilized for these measurements.}
\end{figure}

In the limit where $m_B\rightarrow \infty$ and $x_B\rightarrow 0$, the two axes are separate mass spring systems with well known dynamics. These have frequency response functions (FRFs) that depend on the system parameters $k$, $c$, and $m$ and describe the response of the system to some sinusoidal input. The amplitude and phase of these FRFs are 
\begin{equation}
\frac{X_\mathrm{amp}}{F_\mathrm{amp}/m}(\omega) = \frac{1}{ \sqrt{ (w_0^2 - w^2)^2 + 4 \zeta \omega_0 \omega}}
\label{eq:FRF_amp}
\end{equation}
\noindent and
\begin{equation}
\phi_X - \phi_F = \tan^{-1}\left(\frac{-2 \zeta \omega_0 \omega}{w_0^2 - w^2}\right)
\label{eq:FRF_phase}
\end{equation}
\noindent respectively, where
\begin{equation}
\omega_0 = \sqrt{\frac{k}{m}} \quad \text{ and } \quad \zeta = \frac{c}{2\sqrt{m k}}.
\end{equation}

Figure \ref{fig:frf_meas} shows the Q-band (top) and W-band (bottom) frequency response functions that were measured at the CLASS site, where the drag on the mirror is reduced due to the lower air pressure. The response of the mirror axis to the mirror drive and the reaction axis to the reaction drive shows that the MTM designs are under-damped with resonance frequencies twice (W-band) or below (Q-Band) the drive frequency of 10~Hz. 
\begin{align}
\begin{aligned}
\omega_{0,Q} &< \omega_\mathrm{drive}  \quad &\Delta\phi_Q(\omega_\mathrm{drive}) &\sim -180^\circ \\
\omega_{0,W} &\sim 2 \omega_\mathrm{drive} \quad &\Delta\phi_W(\omega_\mathrm{drive}) &\sim -5^\circ
\end{aligned}
\end{align}

\noindent The resonance frequency of the Q-Band MTM was the main reason for the design changes made to the W-Band MTM (described in section \ref{subsec:W_redesign}). With a phase response of $\sim -180^\circ$, no time-domain controller could produce a stable response at the drive frequency because feedback into the system would be out of phase with the drive. For W-band, a $5^\circ$ phase response means a time-domain feedback control system could be created, however strong filtering must be implemented to prevent the nearby resonance from causing instabilities. In lab tests, the level of filtering required to negate the mechanical resonance of W-band reduced the bandwidth of the control system to below drive frequency. With the significantly reduced bandwidth, the controller was unable to cancel external disturbances and often increased the level of the disturbance through a delayed response. For both MTM designs, a controller which tried to follow a setpoint sinusoidal path in time was not possible or not useful.

Another factor to consider for the control system is that, in practice, $m_B$ is large but finite. The motion of one axis influences the other and external vibrations from forces such as the wind on the telescope transmit through the baseplate. This is visible in figure \ref{fig:frf_meas} where both the mirror and reaction axes have responses to the other axis's drive. There are small features in all the responses which are evidence of resonance frequencies in the support system the VPM was attached to during the measurements. 

Since a time-domain control system was not feasible for either MTM, a frequency-domain control system was developed to drive the axes at a single frequency while ignoring all others. This was possible because observability for the CLASS VPMs is more important than controllability.\footnote{Observability is the ability to know or measure the state of the system while controllability is the ability to move the axes to a particular location in a particular amount of time with some limit on the allowed error. Having sufficient observability of the axis vibrations does require that the output positions are sufficiently sampled to detect them. The VPM positions are read out synchronously with the detectors at 200~Hz which enables the detection of vibrations below the 100~Hz Nyquist frequency. Since the MTM is a mass spring system it functions as a low-pass filter, and significant vibrations above the resonance frequency do not occur.} Demodulation requires knowledge of the positions of the mirror as a function of time; it does not require that those positions be precisely a pre-determined function of time. The use of frequency-domain control does prevent the MTM control from actively reducing external disturbances input on the MTM, meaning all vibration canceling of external impulses is passive and determined by the frequency response function of the MTM.

The MTM controller uses an input amplitude and DC offset for each axis along with a relative phase between the axes to create sinusoidal drive currents for the voice coils. It then observes the resulting axis motion and fits the responses to sine waves of the same frequency to measure the output amplitudes, phases, and offsets from the MTM. This produces a control system that has five inputs, $\vec{X}$, and five outputs, $\vec{I}$,
\begin{align}
\begin{aligned}
\vec{I} &= [I_\mathrm{dc,0},\: I_\mathrm{dc,1},\: I_\mathrm{amp,0},\: I_\mathrm{amp,1},\: \phi_I]^T \\
\vec{X} &= [X_\mathrm{dc,0},\: X_\mathrm{dc,1},\: X_\mathrm{amp,0},\: X_\mathrm{amp,1},\: \phi_X]^T. 
\end{aligned}
\end{align}
\begin{figure}
\centering
\includegraphics[width=0.85\textwidth]{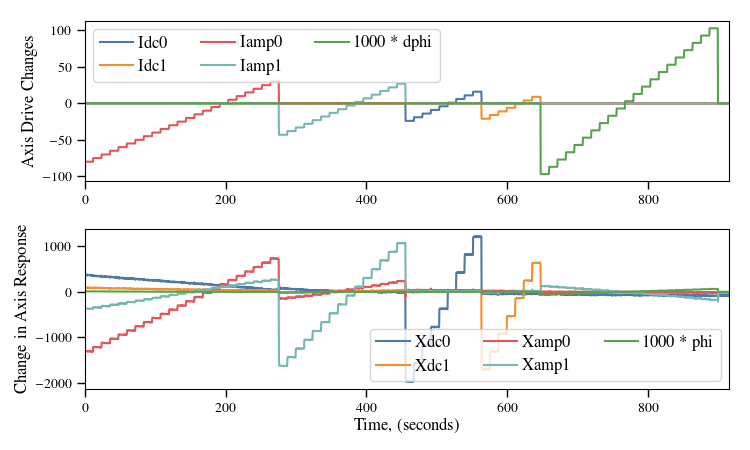}
\caption{\label{fig:covarience_Q} An example of the convenient tuning test used to measure the feedback matrix in equation \ref{eq:feedback_mat}, each of the controller inputs were varied across a range centered on the expected operating point while the response of all the outputs was recorded. This process was performed iteratively to obtain a final value for the feedback matrix used during scientific observations. This example is from the Q-band MTM where the covariance in the response is quite significant.}
\end{figure}
\noindent Assuming this is a linear system, these quantities are related by a system matrix,\footnote{The system matrix described here is not the standard definition of a system matrix used in control theory, where it is usually defined as $\dot{\vec{x}} = A \vec{x} + B\vec{u}$ with $\vec{x}$ being the the state vector and $\vec{u}$ being the inputs to the system.} $\vec{X} = A \vec{I}$. At each step, the change in input current values is calculated as $\Delta\vec{I} = -K (\vec{R} - \vec{X})$, where $K$ is the feedback matrix and $\vec{R}$ is the reference positions defining the VPM throw parameters. In this setup, $K$ is also the inverse of the system matrix, and can be directly measured with the MTM. Figure \ref{fig:covarience_Q} shows an example of the tests used to measure the system matrix for the Q-band MTM, where one input variable was swept across a range of interest and the response of all the output variables was recorded. This type of measurement indicated the system was not completely linear, so subsequent tests were conducted around the set operating points of the MTMs to linearize the system. The feedback matrices of the MTMs can be directly calculated from these tests using 
\begin{align}
K = X_{t}^T I_{t} (I_{t}^T I_{t})^{-1}.
\label{eq:feedback_mat}
\end{align}
\noindent where $I_t$ and $X_t$ are matrices with each row containing the input and output variables of the control system for each time step.

The off-diagonal terms of $K$ are significantly higher for the Q-Band MTM compared to the W-Band MTM, this is likely due to the mechanical dynamics of the MTM and the relation between the drive frequency and the resonance frequency. The off-diagonal terms for Q-Band are large enough that using a diagonal matrix for $K$ is unstable. It is possible to use a diagonal matrix for the W-band MTM; however, the full feedback matrix produces faster settling times and better noise rejection.

\subsubsection{Controller Selection}
As the development of the MTM control system progressed, it was determined that most commercial controllers were not designed for the type of control system required by the MTM. The controllers that were used initially assumed that current was proportional to velocity by default and did not have the available computational capabilities to quickly calculate the output variables for the control system. In particular, the computation rates for fitting the response of the axes limited the bandwidth of the control system to $\sim 0.5$~Hz. These limitations led to the development of custom controllers for the CLASS MTMs built around the OSD3358 system-in-package,\footnote{Octavo Systems, \url{https://octavosystems.com/}} which includes a 1\,GHz ARM Cortex-A8 processor and 512\,MiB of RAM, providing ample computational resources to run the control code using a real-time operating system. The chip also includes dedicated hardware peripherals for reading in the MTM's quadrature encoders and an Ethernet interface for command input and telemetry. To actuate the voice coils, the controller includes a pair of H-bridge coil drivers with current sensing capabilities. A high sensitivity 3-axis accelerometer is used for tuning the reaction canceling axis, a process that is described in section \ref{subsec:vib}. The new controllers increased the control bandwidth to $\sim2.5$~Hz and this bandwidth is no longer limited by the controller capabilities.

\subsection{\label{subsec:vib} Vibration Suppression}

\begin{figure}
\centering
\begin{tabular}{cc}
\includegraphics[width=0.45\textwidth]{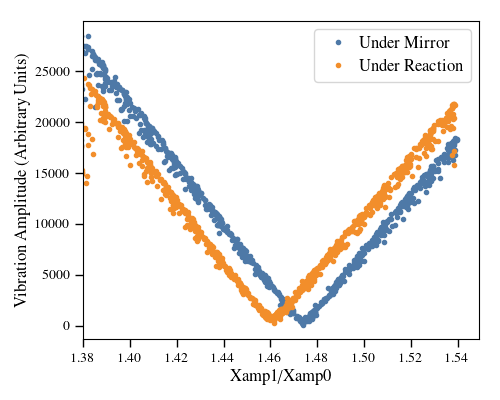} & 
\includegraphics[width=0.45\textwidth]{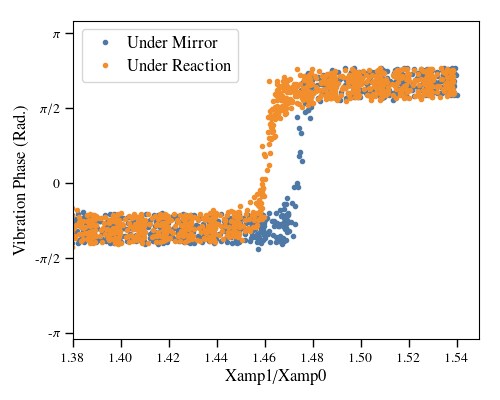}\\
\end{tabular}
\caption{\label{fig:accel_cancel} The amplitude and phase (relative to the mirror axis position) of the vibrations measured for the Q-band VPM with the accelerometer positioned at different places on the MTM flexure baseplate. These measurements indicate that the vibrations can be significantly canceled using the correct setting for the reaction axis relative to the mirror axis and that the placement of the accelerometer during this tuning is important for canceling vibrations at the correct locations within the telescope structure.}
\end{figure}

The impulses created by the mirror motion are significant and must be canceled to avoid unnecessary stress on the telescope structure. This is accomplished with the reaction canceling axis of the MTM. The control settings for the reaction axis determine the level of vibration canceling and must be correctly tuned.

Looking back at figure \ref{fig:q_v_w_control} and assuming this simple model of the mechanical setup, the vibrations from the MTM will be zero when $\ddot{x}_B = 0$. If the mirror and reaction axes are identical ($k_0=k_1$ and $c_0=c_1$) and moving at 10~Hz with amplitudes of $X_\mathrm{amp,0}$ and $X_\mathrm{amp,1}$ respectively, then $\ddot{x}_B = 0$ when 
\begin{equation}
\frac{X_\mathrm{amp,1}}{X_\mathrm{amp,0}} = \frac{m_0}{m_1}.
\end{equation}
\noindent In practice, the values for $m_i$ are related but not equal to the physical mass of the axes and $c_0 \neq c_1$ because of the mirror's larger surface area and the effect of the reaction axis counterweight. In addition, the MTM has a finite size and vibrations have finite wavelengths, meaning the level of vibration canceling will vary across the MTM. Despite these complications, it is still possible to find a ratio $X_\mathrm{amp,1}/X_\mathrm{amp,0}$ which minimizes the vibrations at locations of interest. For example, figure \ref{fig:accel_cancel} shows the level of vibrations measured along the axis of motion with accelerometers under either the mirror or reaction axis voice coils for a range of $X_\mathrm{amp,1}/X_\mathrm{amp,0}$ values. These measurements were performed on the Q-Band MTM before the wire grid was installed. The clear minimum and simultaneous phase shift distinctly show the optimal ratio to cancel vibrations at each location. These measurements were completed before and after the installation of the wire grid as well as once the VPM was installed in the telescope structure. Reoptimization is critical after installation because the cage mounts are much sturdier than those used during MTM development and testing.

After the Q-band VPM was installed on the telescope mount, accelerometers and the mount encoders were used to improve the reaction canceling within the complete telescope system. The residual vibrations from the VPM in the telescope structure are well below other vibrational lines in the system.

\section{Wire Grids}
\label{sec:wiregrid}

\begin{figure}
\begin{floatrow}
\floatbox[\nocapbeside]{table}[\FBwidth][\FBheight][l]
{\caption{\label{tab:gridspecs}Properties of the CLASS wire grids}}
{  \begin{tabular}{l|c} 
  Grid property & Value \\ \hline\hline
  Wire material  &  Tungsten\\
  Wire plating & Copper (5 $\mu$m)\\ 
  Wire diameter & 50 $\mu$m\\
  Wire spacing & 160 $\mu$m\\
  Grid diameter & 62 cm\\\hline
  \end{tabular}
}
\floatbox[\nocapbeside]{figure}[\FBwidth][\FBheight][c]
{\includegraphics[width=0.5\textwidth]{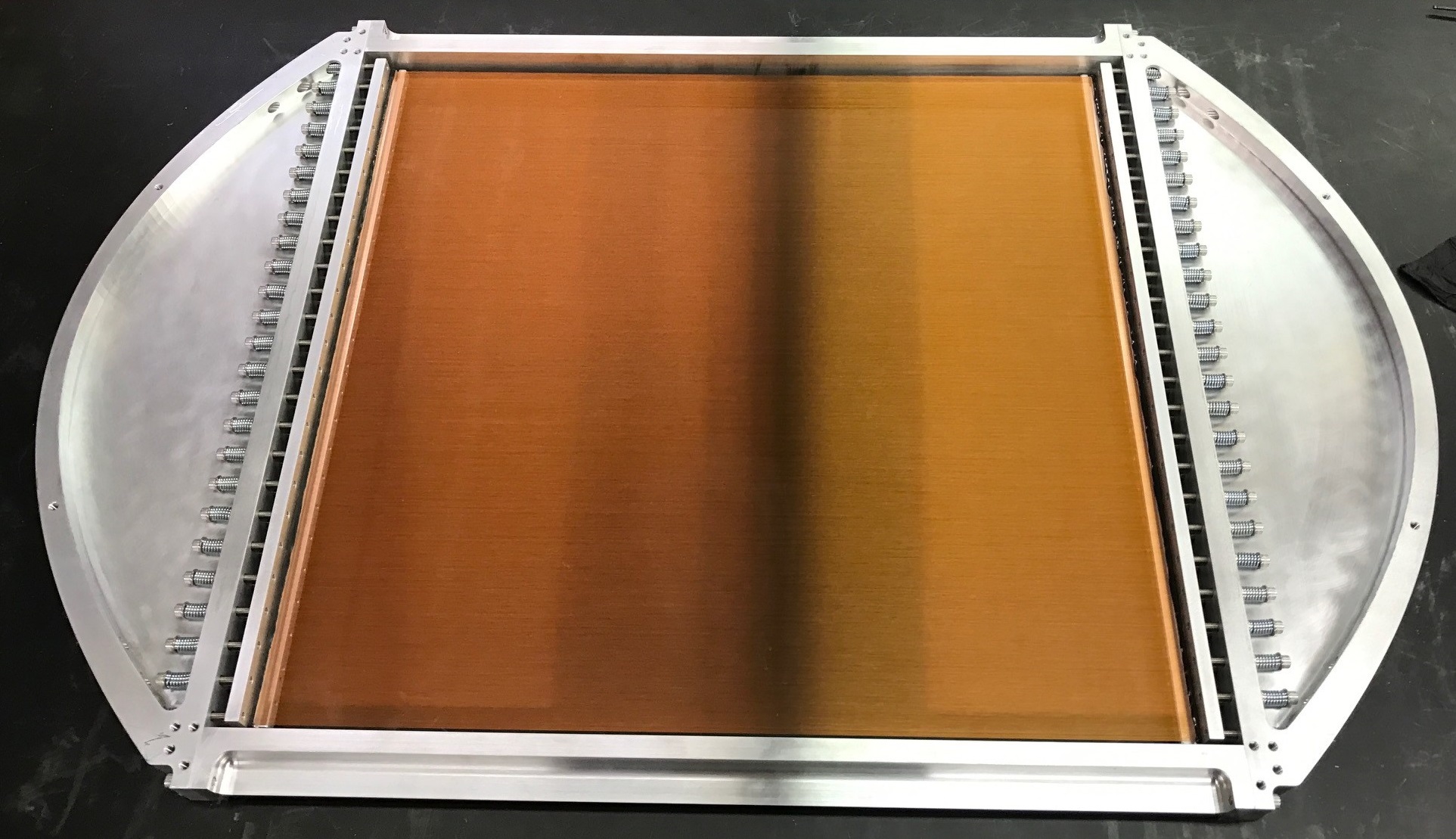}}
{\caption{\label{fig:wire_grid}  The VPM grid for the first W-Band CLASS telescope}}
\end{floatrow}
\end{figure}

A stationary linearly polarizing wire grid is held $\sim1$~mm in front of the moving mirror and is used to separate the two orthogonal linear polarization states of the incoming radiation. Real wire grids have several non-ideal effects which can be mitigated though optimization of the materials and specifications of their design. Reference \citenum{Chuss2012} developed a transmission-line optical model of VPMs at submillimeter wavelengths and showed that a ratio of $\pi$ between the wire pitch ($g$) and wire diameter ($2a$) is a reasonable choice for producing grids with high isolation between orthogonal linear polarizations but that small variations around this ratio did not have a significant impact on the isolation. In addition, the ratios of $g/\lambda$ and $a/\lambda$ should be made as small as possible, and the outer material of the wires should be as conductive as possible in order to minimize loss and thermal emission from the wires.

CLASS chose to use copper plated wires because measurements in reference \citenum{Chuss2012} indicated a gold wire plating might not function optically as gold. Copper plating will develop a copper oxide layer, but it will be extremely subwavelength. Also the copper plating can be made several times thicker than the skin depth at 40~GHz. Tungsten is used as the bulk material for the wires because it has the highest yield strength to density ratio of standard materials that can be drawn of formed into small caliber wire. The diameter of the CLASS wires was chosen based on how the wires would be secured to the grid frame. Stycast\textsuperscript{\textregistered} 2850 with 23LV catalyst was used to epoxy the wires onto the grid mandrel bars. Pull tests with a variety of diameters indicated that the 50~$\mu$m wire would be secured by the epoxy up to its breaking strength of $\sim 7$~N while 38~$\mu$m wire would pull out of the epoxy before it broke.  Wires smaller than 50~$\mu$m could be used with other techniques to secure them, such as those used for the PIPER grids\cite{Chuss2014}; however, 50~$\mu$m is sufficiently subwavelength for the CLASS VPMs to allow this simpler method of wire attachment.

The wire grids for the VPMs for the four CLASS telescopes were manufactured using a wire wrapping technique first pioneered in the far-infrared \cite{Novak1989} and later scaled to larger format \cite{Voellmer2008}. The details of the CLASS grid wrapping technique are similar to those used for the PIPER grids \cite{Chuss2014}.  In this process, the wires are wrapped onto a 8-inch diameter mandrel. Two aluminum bars are counter-sunk into the mandrel and grooved at the desired wire spacing using a square straight flute carbide end mill held at $45^\circ$ relative to the mandrel. Wires are wrapped using a CNC mill with a 4th (rotation) axis mounted to the machine bed.  After wrapping is complete, the wire is epoxied to the mandrel bars. The wire between the mandrel bars is cut, and the wires are carefully unwrapped and mounted to a frame using 100 screws and springs. The screws are tightened a set amount such as to apply 50\% of the breaking strength to each wire which requires over 3000~lbs of force across the bars holding the wires. The applied tension raises the resonance frequency of the wires to over an order of magnitude above the modulation frequency. After over a year of operation at the CLASS site, all of the wires comprising the Q-band wire grid were still intact. Figure \ref{fig:wire_grid} shows the complete grid for the first W-band VPM.

\begin{figure}
\centering
\begin{tabular}{cc}
\includegraphics[width=0.33\textwidth , align=c]{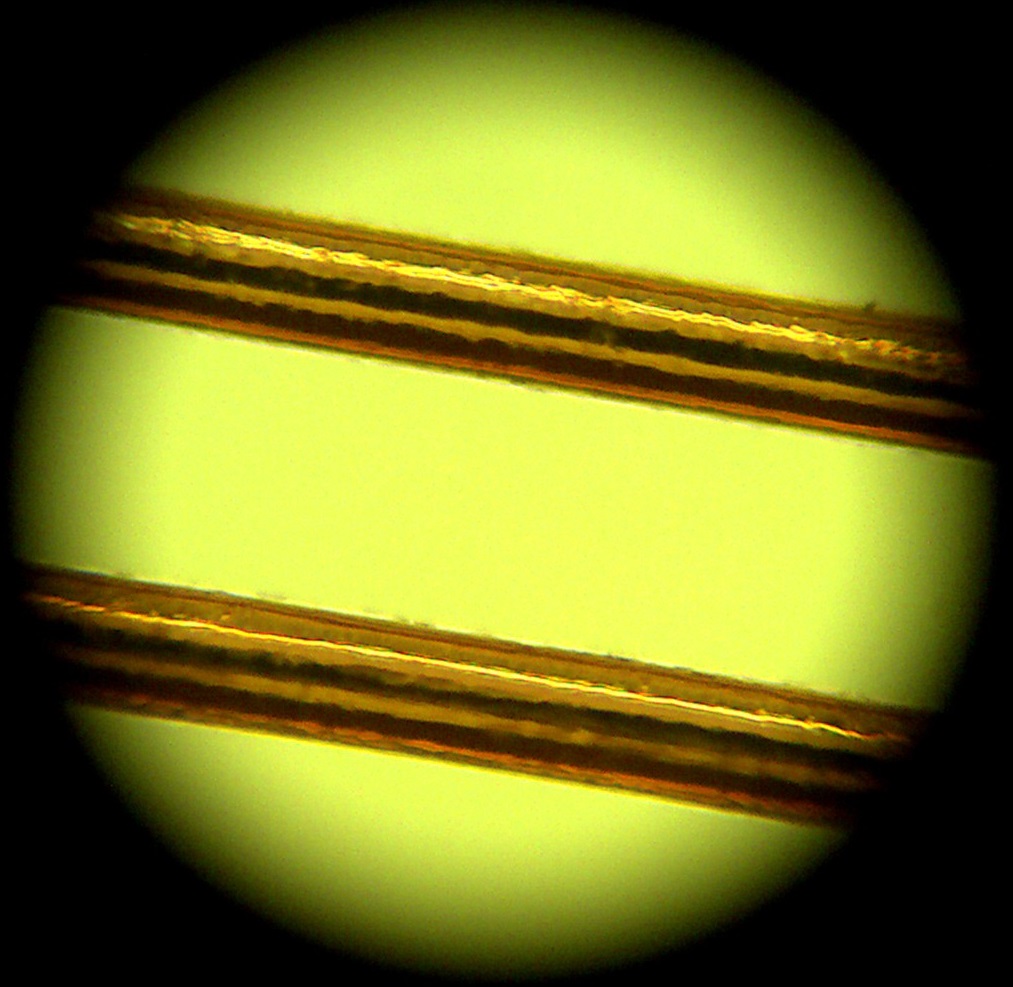} &
\includegraphics[width=0.6\textwidth , align=c]{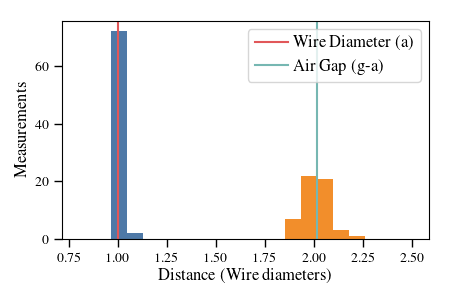} \\
\end{tabular}
\caption{\label{fig:wire_pitch}  (Left) A microscope image of the wires taken using the XY-gantry calibration microscope, used to measure the wire spacing across the grid and to align the grid plane to the mirror. (Right) The wire spacing measurements for the W-band wire grid. Fifty images of the wires were taken across the face of the grid. The diameters of the wires in those images were used to calibrate the spacing between the wires and determine the pitch of the grid.}
\end{figure}

Figure \ref{fig:wire_pitch} shows a microscope image taken using a microscope mounted on a custom XY-gantry system described in section \ref{sec:alignment}. Fifty of these images were taken across the vertical height of the grid ($\sim1$ per 0.5~in) with the microscope centered horizontally. These images were used to measure the ratio of the wire diameter to wire spacing. For the W1 grid the pitch to diameter ratio is $3.01\pm 0.06$, which is within $5\%$ of the target spacing of $\pi$ and a histogram of the measured values is shown in the right plot of figure \ref{fig:wire_pitch}. A similar procedure was done for the Q-band wire grid which has a pitch to diameter ratio of $2.61\pm 0.12$, which is $17\%$ away from the target spacing but still within the acceptable range for the CLASS grids.

\section{Grid-Mirror Alignment}
\label{sec:alignment}

The wire grids need to be held flat and parallel to the VPM mirror in order to reduce temperature to polarization leakage and maintain the polarization transfer function as close to ideal as possible. In addition, the absolute position of the wire grid with respect to the mirror-off position\footnote{The mirror-off position or the zero current position is defined as the position the MTM sags to when the telescope is pointed at $45^\circ$ elevation and the voice coils are unpowered.} has to be within 2~mm (Q-band) or 0.2~mm (W-band) of the desired throw center. This requirement is determined by the stiffness of the MTM flexure, the weight of the mirror, and the force available from the voice coils. 

\begin{figure}
\begin{tabular}{cc}
\includegraphics[width=0.45\textwidth]{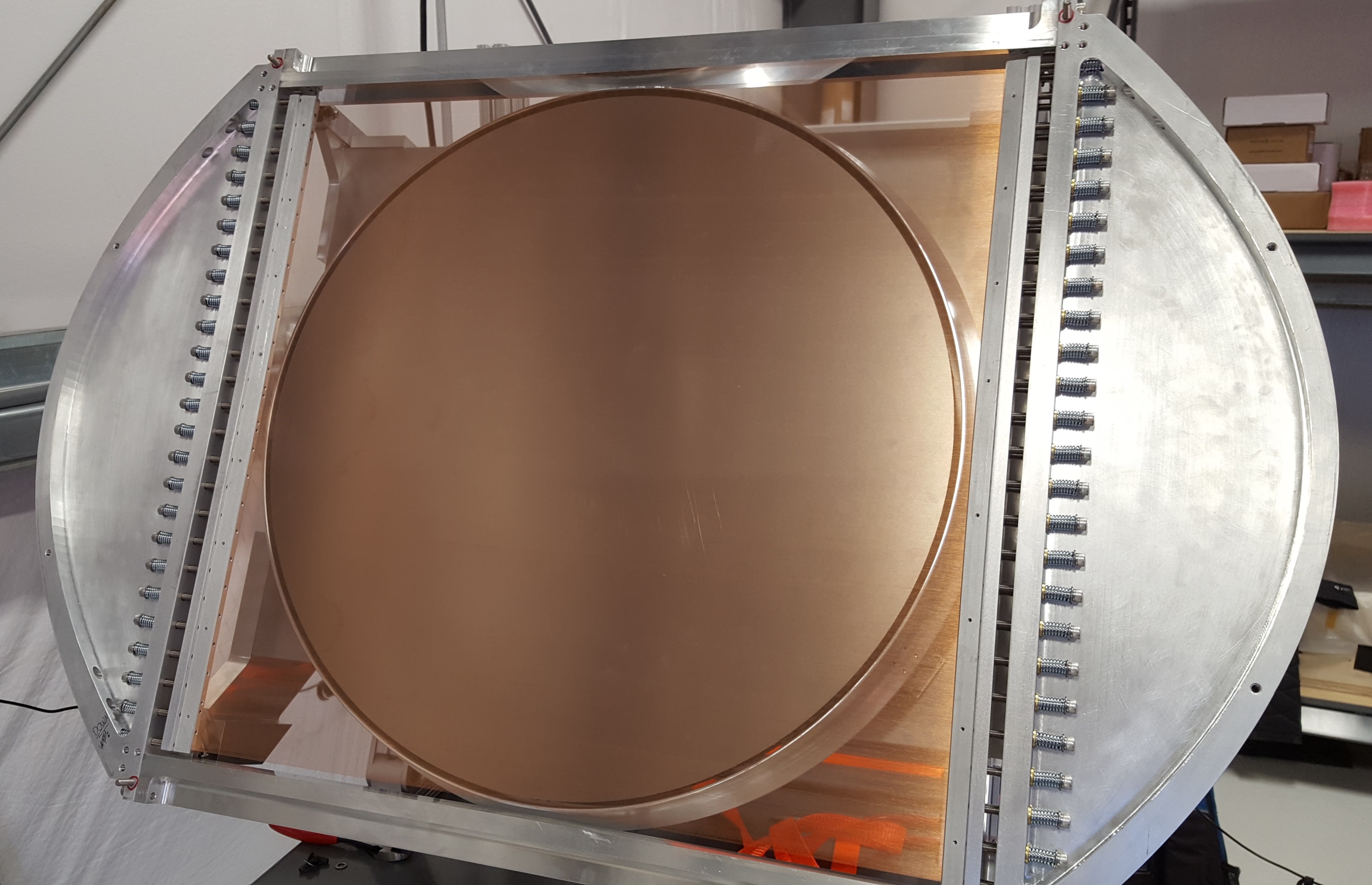} & 
\includegraphics[width=0.50\textwidth]{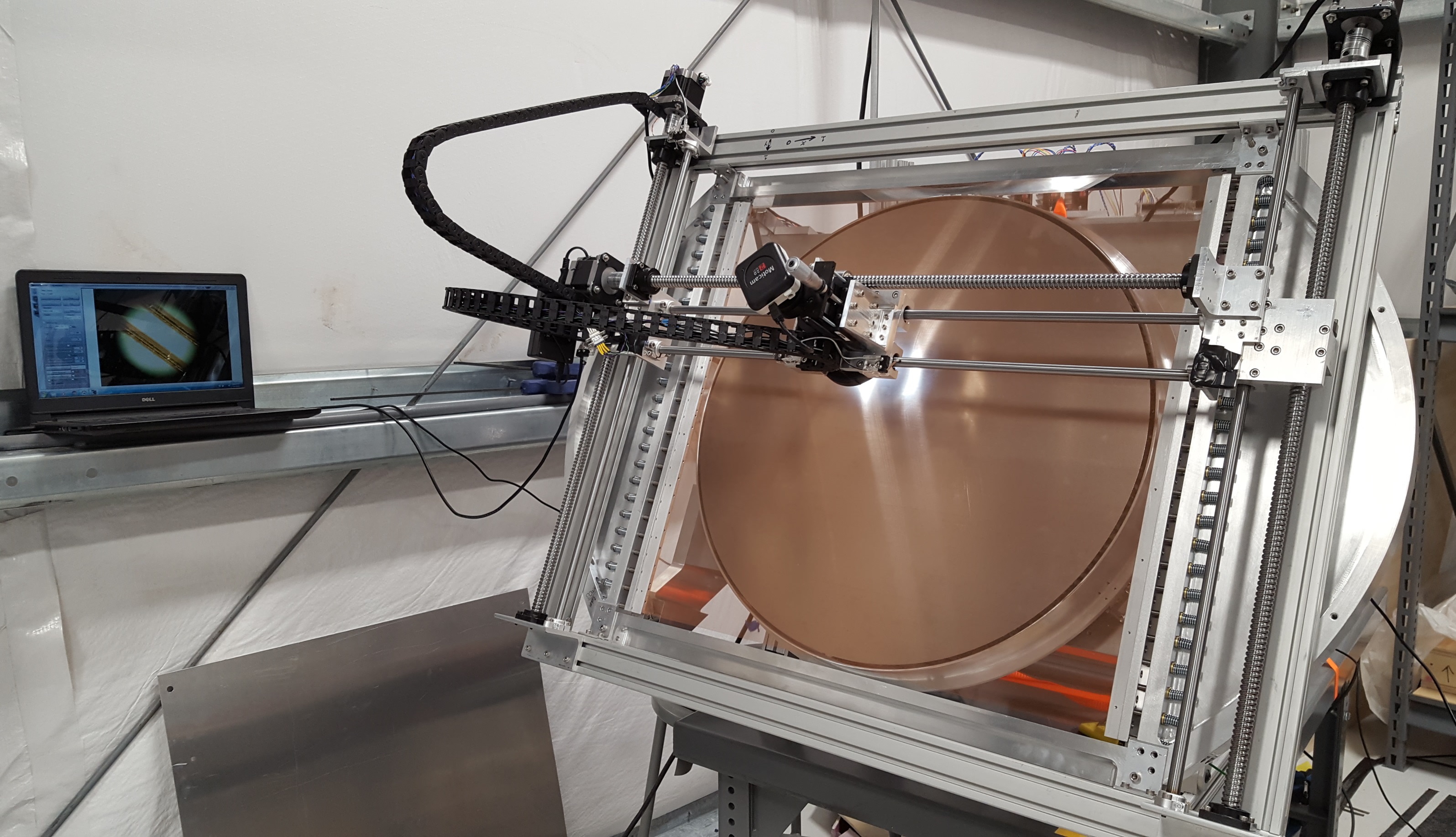} \\
\end{tabular}
\caption{\label{fig:w1_vpm} Left: The completed W-band VPM in the highbay at the CLASS site. Right: The XY-gantry calibration microscope in used on the W-band VPM.}
\end{figure}

To complete this alignment, a custom XY-gantry calibration microscope was designed and built because no commercial XY-gantry could meet the requirements of travel and clearance across the grid face. The left image in figure \ref{fig:w1_vpm} shows the XY-gantry in use with the W1 VPM. Three 750~mm long 16~mm diameter ball screws, two on the vertical axis and one on the horizontal axis, are coupled to NEMA~23 stepper motors. The ball screws and four 12~mm diameter linear shafts support the microscope stage which has 546~mm by 570~mm of travel. The microscope stage contains a microscope connected to a vertically mounted linear micrometer stage with 25~mm of travel. The microscope has a 10x objective and 20x wide-field eyepiece for 200x total magnification and is coupled to a 2 megapixel USB camera through a macro lens. The image of the wires in figure \ref{fig:wire_pitch} is an example of the view of the wires through the microscope, which has a $\sim250$~$\mu$m diameter field of view. The microscope is focused on the wires or the mirror using the micrometer stage on the gantry. The depth of focus is $\sim10$~$\mu$m, which is small enough to allow the user to distinguish between the top and edge of each wire. The positioning of the microscope stage is controlled with an Arduino Uno\footnote{\url{https://www.arduino.cc/}} connected to a SMAKN\textsuperscript{\textregistered} stepper motor driver. In use, the microscope stage can be repeatably stepped the $\sim 160$~$\mu$m between wires, and reliably returns within $\sim 50$~$\mu$m to previous positions after removing the gantry from the VPM and reinstalling.

\begin{figure}
\begin{tabular}{c}
\includegraphics[width=0.95\textwidth]{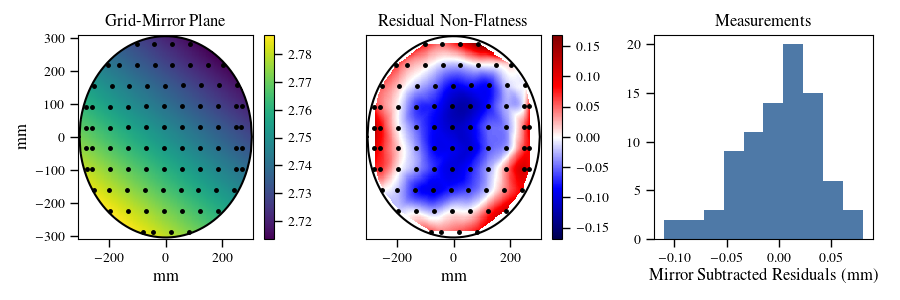} \\
\includegraphics[width=0.95\textwidth]{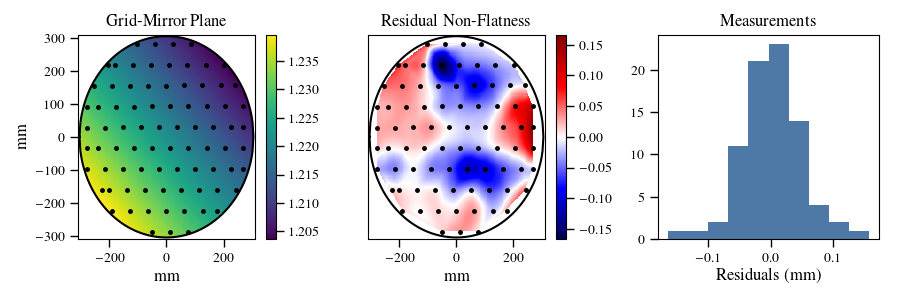} 
\end{tabular}
\caption{\label{fig:gm_alignment} The final alignment of the grid to the mirror for the Q-band (top) and W-band (bottom) VPMs. The left figures show the plane of the grid relative to the mirror at the mirror-off position and the black dots indicate the measurement points. The center plots show the residuals from the plane fit and, for the Q-band, primarily shows the shape of the mirror described in section \ref{subsec:mirror}. The colorbar scales of the left and center plots are all in millimeters. The right plots show the histograms of the residuals to the plane fit, the Q-band residuals have also had the CMM measured mirror shape was subtracted.}
\end{figure}

To align the wire grid to the mirror, the distance between the wires and the mirror is measured at $\sim60$ points across the mirror face by measuring the distance the vertical linear stage must be adjusted to change the focus between the wires and the mirror. This distance is recorded with the XY-gantry position and the three mirror encoder positions to determine the plane of the wire grid with respect to the mirror plane at the mirror-off position. The parallelism and positioning of the wire grid is coarsely adjusted with spacers under the kinematic grid mounts and then finely adjusted with micrometers attached to a flattening ring that is gently pressed into the wires to define a wire plane. An initial alignment was performed at JHU and the final alignments were performed at the CLASS high-altitude site before the VPMs were installed in the telescope optics cage.

The top row in figure \ref{fig:gm_alignment} shows the final alignment for the Q-band VPM. The top left plot shows a plane fit to the grid-mirror distance measurements across the face of the mirror. The center plot shows the residuals to that fit. The shape in the residual non-flatness is due to the shape of the Q VPM mirror described in section \ref{subsec:mirror}. The tilt across the mirror is 24.9 arcseconds, which is 1.5 wire diameters and well within the tolerances required for the 90 arcminute Q-band beam. The residuals from the measurement, after the CMM measured mirror shape is subtracted, indicates the wire grid has an RMS flatness error of 36~$\mu$m with the flattening ring engaged, which is less than the diameter of a wire.

The bottom row in figure \ref{fig:gm_alignment} shows the final alignment for the first W-band VPM. The top left plot shows a plane fit to the grid-mirror distance measurements across the face of the mirror. The center plot shows the residuals to that fit. The tilt across the mirror is 12.2 arcseconds, which is half the diameter of a wire and well within the tolerances required for the 40~arcminute W-band beam. The residuals from the measurement indicate the wire grid has an RMS flatness error of 49~$\mu$m when the flattening ring is engaged, which is about the diameter of one wire.

\begin{figure}
\centering
\begin{tabular}{cc}
	\includegraphics[height=1.9in]{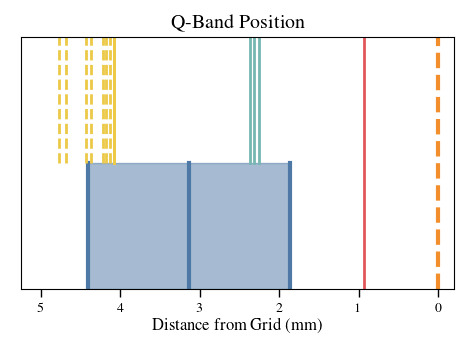} &
	\includegraphics[height=1.9in]{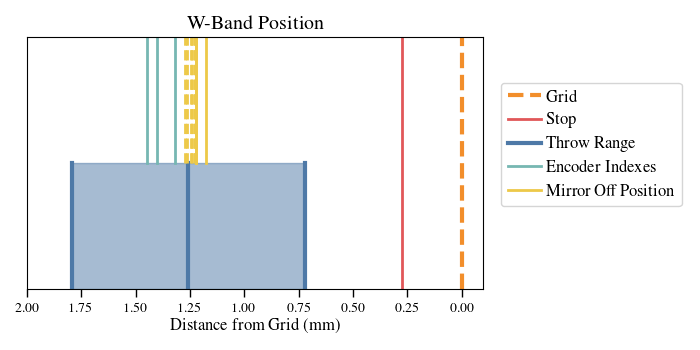} \\
\end{tabular}
\caption{\label{fig:vpm_positions} The absolute position of the Q- and W-band VPM mirrors with respect to the wire grid after alignment. The mirror-off positions (plotted for each boresight value) are within the required range for each VPM and the encoder index locations are within the mirror throw. The red line marks the location of a hard mirror stop that is a protection of last resort for the VPM grid.}
\end{figure}

In addition to the parallelism requirements, the absolute position of the wire grid with respect to the mirror-off position and the encoder indexes needed to be set. This requirement was much more stringent for the W-Band VPM as the stiffer flexure made the total possible displacement of the mirror much smaller. Figure \ref{fig:vpm_positions} shows the achieved Q-band and W-band positions. The red lines indicate the position of the hard stop for the VPMs, where a metal block is positioned to prevent farther travel. The distances between the mirror-off positions and the required mirror throw are within the range accessible to each VPM, and the encoder indexes are within the mirror throw, enabling position recalibration during observation.

\begin{figure}
\centering
\begin{tabular}{cc}
	\includegraphics[width=0.45\textwidth]{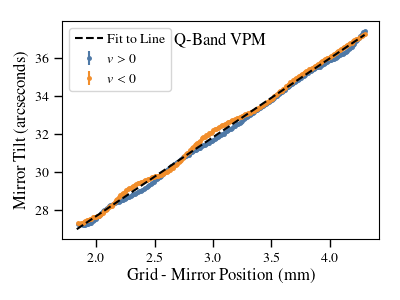} &
	\includegraphics[width=0.45\textwidth]{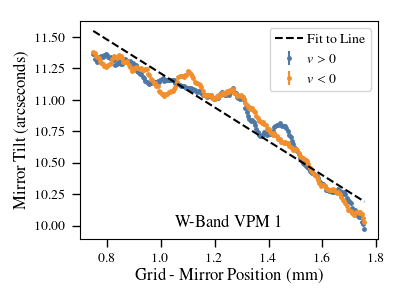} \\
\end{tabular}
\caption{\label{fig:parallelism_over_throw}  The tilt or parallelism error between the mirror and grid plane as the VPM is running at 10~Hz at an elevation $=45^\circ$ and boresight $=0^\circ$ for the Q-band (left) and W-band (right) VPMs. The absolute value of these tilts are determined with the grid-mirror alignment and then relative tilts are calculated from each set of encoder positions during the test.}
\end{figure}

\section{VPM Performance}

The tilt of the VPM mirrors is measured using three optical linear encoders mounted on the back of the mirror. The change in tilt across the throw of the VPM creates a temperature to polarization leakage in single detector data. While various analysis methods, such as template subtraction, can be used to measure and remove these signals, the design goal of the MTM was to reduce their amplitude as much as possible. 

Figure \ref{fig:parallelism_over_throw} shows the amplitude of the mirror tilt as a function of grid mirror distance while the VPMs were running at 10~Hz for the Q-Band (left) and W-Band (right) VPMs. Both measurements were taken with the VPM held in the configuration consistent with the CLASS telescope observing at $45^\circ$ elevation and at zero boresight. The absolute value of the tilt was calibrated using the grid-mirror alignment, described in section \ref{sec:alignment}, which results in a parallelism measurement relative to a plane defined by a set of three encoder values. The tilts are then calculated for each set of encoder values as the VPM runs. Fitting the tilts to lines gives a changing tilt of $4.16$~arcsec/mm and $-1.35$~arcsec/mm for the Q-band and W-band respectively; however, it is clear that a linear fit is not a complete explanation. The overall reduction in the tilt slope between the Q-band and W-band is likely due to the increased stiffness in the MTM flexure. 

Both the Q-band and W-Band tilts exhibit a hysteresis effect from the direction of motion of the mirror. The effect is $\pm 0.3$~arcseconds for Q-band and $\pm 0.15$~arcseconds for W-band. This is most likely due to the air resistance on the mirror motion because it depends on velocity and the maximum velocity for Q-band is twice that of W-band. The air resistance could produce a changing force vector on the mirror supports that could cause a slight tilt. It is possible the other non-linear aspects of the tilts come from process variation in the assembly of the CLRFs or small machining errors in the linkages of the full flexure.

The range of the Q-band tilt is 10.25~arcseconds which indicates a level of $T\rightarrow P \sim 3\times10^{-5}$ for single detector data. The tilt of the W-band mirror over the mirror throw, 1.4~arcseconds, implies $T\rightarrow P \sim 3\times10^{-6}$. With forward modeling or pair subtraction, the $T\rightarrow P$ due to tilts can be further reduced. In addition, the front-end modulation of the CLASS data with the VPM means $T\rightarrow P$ effects at or before the VPM, such as with these tilts, are some of the few sources of instrument polarization that can affect CLASS data.

\begin{figure}
\centering
\includegraphics[width=0.9\textwidth]{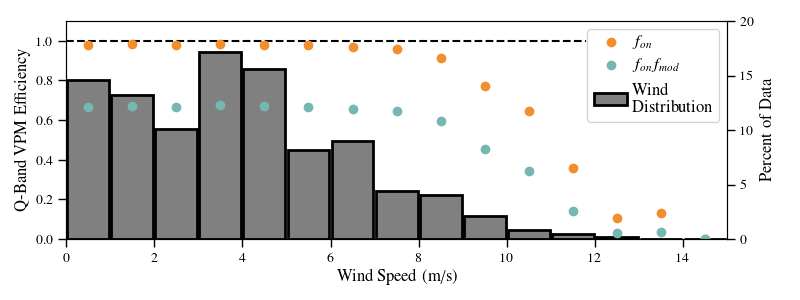}
\caption{\label{fig:q_vs_wind} The VPM on efficiency and the VPM modulation efficiency as a function of wind speed for the Q-band VPM between September 1st 2016 and February 1st 2018. The bars denote the distribution of the winds experienced at the CLASS site during this time.  The low resonance frequency of the Q-band MTM limits the VPM functionality in high winds, but the overall on efficiency is still over 96\%.}
\end{figure}

The Q-Band VPM was installed at the CLASS site in the Spring of 2016. As described in section \ref{subsec:W_redesign}, the lower than ideal resonance frequency of the Q-band MTM allows the wind and motion of the telescope to excite vibrations in the mirror. These vibrations do not decrease the quality of data obtained by the experiment because the exact mirror positions are read out fast enough to observe the vibrations and these positions are used in the demodulation of the CLASS data. The vibrations do inhibit the function of the VPM in high winds ($\geq 12$~m/s) because the mirror will hit the software limit and the MTM control will turn off to protect the wire grid. To mitigate this effect, the total throw of the VPM was reduced slightly to allow for more vibrations. This reduced the overall modulation efficiency but significantly increased the uptime of the VPM. 

Figure \ref{fig:q_vs_wind} shows the effect of the wind on the uptime of the Q-band VPM at the CLASS site between September 1st 2016 and February 1st 2018. During this time period the VPM was on for over 288 days and completed over 250 million cycles. This accounts for over 96\% of the time the Q-band telescope spent observing the CMB. The VPM modulation efficiency, defined in equation \ref{eq:fmod}, during this period was 0.675 which is lower than the ideal value but was required to maximize the efficiency of the survey.

\section{Conclusions}
The VPMs for the CLASS telescopes were designed to meet a variety of constraints that were required to achieve the science goals of the CLASS survey. The CLASS VPMs required a 60~cm clear aperture with a mirror throw of up to 2.5~mm at 10~Hz while maintaining excellent parallelism with between the mirror and the wire grid. The flexure-based mirror transport mechanism for the Q-band and first W-band VPMs achieve these requirements; however, design optimization was required between the Q- and W-band MTMs. The VPMs for the second W-band and the high frequency CLASS telescopes have the same design as the first W-band VPM.

The four wire-grids for the CLASS VPMs have been fabricated, and the specifications for the first two to be used are reported here. These large aperture wire grids have unprecedented uniformity and flatness across the entire VPM aperture and the manufacturing technique employed here is scalable to even larger apertures.

\section*{Acknowledgments}
We acknowledge the National Science Foundation Division of Astronomical Sciences for their support of CLASS under Grant Numbers 0959349, 1429236, 1636634, and 1654494. The CLASS project employs detector technology developed under several previous and ongoing NASA grants. Detector development work at JHU was funded by NASA grant number NNX14AB76A. K. Harrington is supported by NASA Space Technology Research Fellowship grant number NX14AM49H. T. Essinger-Hileman was supported by an NSF Astronomy and Astrophysics Postdoctoral Fellowship. We further acknowledge the very generous support of Jim and Heather Murren (JHU A\&S ’88), Matthew Polk (JHU A\&S Physics BS ’71), David Nicholson, and Michael Bloomberg (JHU Engineering ’64). CLASS is located in the Parque Astronómica Atacama in northern Chile under the auspices of the Comisión Nacional de Investigación Científica y Tecnológica de Chile (CONICYT).

\bibliography{class_bib}   
\bibliographystyle{spiebib}   

\end{document}